\documentclass[aps,pra,epsf,superscriptaddress,amsmath,amssymb,amsfonts,twocolumn,showpacs,nofootinbib]{revtex4-1}

\usepackage{graphicx}
\usepackage{dcolumn}
\usepackage{bm}
\usepackage{braket}
\usepackage{amsmath}
\usepackage{mathtools}
\usepackage{graphicx,color,xcolor}
\usepackage{hyperref}
\usepackage[capitalize]{cleveref}
\newcommand{\abs}[1]{\left| #1 \right|} 
\usepackage{multirow}

\usepackage[normalem]{ulem} 

\usepackage{xcolor}
\usepackage{hyperref}
\hypersetup{
    colorlinks=true,
    linkcolor=cyan,
    urlcolor=cyan,
    citecolor=cyan
}

\begin{document}

\title{Localization and splitting of a quantum droplet with a potential defect}

\author{F. Bristy}
\affiliation{Department of Physics and LAMOR, Missouri University of Science and Technology, Rolla, MO 65409, USA}

\author{G. A. Bougas}
\affiliation{Department of Physics and LAMOR, Missouri University of Science and Technology, Rolla, MO 65409, USA}

\author{G. C. Katsimiga}
\affiliation{Department of Physics and LAMOR, Missouri University of Science and Technology, Rolla, MO 65409, USA}%

\author{S. I. Mistakidis}
\affiliation{Department of Physics and LAMOR, Missouri University of Science and Technology, Rolla, MO 65409, USA}

\date{\today}

\begin{abstract}
We unravel the existence and nonequilibrium response of one-dimensional harmonically trapped droplet configurations in the presence of a central 
potential barrier or well.
For fixed negative chemical potentials, it is shown that droplets fragment into two for increasing potential barrier heights, a process that occurs faster for larger widths. 
However, atoms from the droplet accumulate at the potential well, especially for wider ones, leading to a deformed droplet and eventually to the termination of the solution. Linearization analysis yields the underlying excitation spectrum which dictates stability and the behavior of the ensuing collective modes. Quenches in the potential height are used to demonstrate dynamical fragmentation of the droplet for potential barriers as well as self-evaporation along with droplet localization and eventual  relaxation for longer evolution times in the case of 
potential wells. 
The presence of selective excitation processes emanating from quantum superposition in  
the induced droplet dynamics is explicated by evaluating the contribution of the participating single-particle eigenstates.  
Our results should be detectable by current ultracold atom experiments and may inspire engineered droplet dynamics with the aid of external potentials. 
\end{abstract}

\maketitle

\section{Introduction}

Recent ultracold atom experiments offer access to the regime of weak attractive interactions~\cite{bottcher2020new,chomaz2022dipolar,Chen_observation_2021}. 
Such settings can host, among others, ultradilute liquid-type many-body droplet states~\cite{luo2021new,malomed2020family,mistakidis2023few} that are sustained due to the presence of quantum fluctuations in dipolar gases~\cite{bottcher2020new,politi2022interspecies,chomaz2022dipolar}  as well as homonuclear~\cite{cheiney2018bright,semeghini2018self,cabrera2018quantum} and heteronuclear~\cite{d2019observation,burchianti2020dual} short-range Bose mixtures.   
Commonly the interaction~\cite{bottcher2020new} 
and dimension~\cite{Ilg_crossover_2018,pelayo2025phases} dependent Lee-Huang-Yang (LHY) first-order correction term~\cite{lee1957eigenvalues} to the mean-field energy suffices to capture the formation and dynamics of quantum droplets within a suitable extended Gross-Pitaevskii (eGPE) model~\cite{petrov2015quantum,petrov2016ultradilute,englezos2025multicomponent}. 
Beyond LHY effects have been, for instance, documented in Refs.~\cite{parisi2019liquid,mistakidis2021formation,englezos2023correlated}. 

Droplets in bosonic mixtures exhibit a constant flat-top density~\cite{tylutki2020collective} for increasing atom number manifesting their inherent surface tension and incompressibility~\cite{petrov2015quantum,luo2021new,Ancilotto_tension}. They may also assemble in mixed droplet-gas configurations under intracomponent interaction or particle imbalance~\cite{Flynn_harm,pelayo2025droplet}.  
Droplets are many-body self-bound states, that 
can break apart when subject to large perturbations (self-evaporation)~\cite{Ferioli_evaporation,fort2021self}, and suffer three-body losses~\cite{semeghini2018self} obstructing their long time observation.       
Several properties of these exotic phases-of-matter have been recently unveiled
such as their capability to accommodate nonlinear solitary waves~\cite{edmonds2023dark,katsimiga2023solitary} and vortical patterns~\cite{tengstrand2019rotating,li2018two,Bougas_vortex_drops}, prevent transverse destabilization of kink structures~\cite{Mistakidis_kinkdrop}, quasi-elastic collisional features~\cite{ferioli2019collisions} and triggering of modulational instability~\cite{mithun2020modulational,otajonov2022modulational}.

On the other hand, the immersion of impurity atoms into droplets can bring forth additional mechanisms including self-localization of the impurity~\cite{abdullaev2020bosonic,Bighin_localization}, hybrid droplet-impurity modes~\cite{Sinha_imp_drop}, tunability of the dipolar bound state character~\cite{wenzel2018fermionic}, and mediated impurity interactions from the droplet~\cite{pelayo_BF_drops}. 
The impact of a heavy impurity on its host bosonic droplet remains, however, elusive. 
As demonstrated in repulsive Bose gases~\cite{Campbell,brauneis2022artificial}, the heavy impurity can be 
well approximated by an external potential barrier, an approach we also adopt in this work.
In this context, a repulsive barrier dragged through the droplet has been used recently to dynamically generate defects like solitary waves and vortices~\cite{Saqlain,dos2025supersonic}, a well-established protocol in repulsive Bose gases~\cite{Engels_obstacle,Hakim,Rodrigues,Katsimiga_obst} but with an amended local speed of sound. 
Additionally, the tunneling dynamics of droplets in a double-well potential have been discussed~\cite{Abdullaev_DW,Wysocki_DW} as well as the transmission and reflection properties of droplets passing through a 
potential well have been reported~\cite{debnath2023interaction}.

Accordingly, it is interesting to study the effect of the impurity onto the droplet such as to retain its bound state character in a structurally modified configuration or its destruction. 
Here, the computation of the underlying excitation spectrum being able to reveal the stability of the resulting states and their ensuing modes is certainly desirable. 
In another vein, the dynamical response of the droplet by switching-on the impurity can trigger specific excitation modes or even fragmentation. 
To explore these possibilities, we deploy either a potential barrier or well in a symmetric~\cite{petrov2016ultradilute} (alias single-component) one-dimensional (1D) quantum droplet featuring contact interactions. Notice that in the limiting case of a delta-like potential barrier (well) the composite system resembles an infinitely heavy impurity pinned at the trap center interacting repulsively (attractively) with its droplet host, while the impurity-droplet coupling is regulated by the value of the potential height.
Similar external confinements with attractive single-particle potentials are used to emulate the motion of electrons in solids~\cite{atkins2011molecular}, which further promotes the broader character of our results. 
The static 
properties and nonequilibrium dynamics of this setting are modeled by the appropriate eGPE~\cite{petrov2016ultradilute}.

On the ground-state, we determine regions of existence of droplet states in the  presence of a potential well or barrier. 
The former favors droplet's spatial localization as was  also argued in Ref.~\cite{debnath2023interaction}. 
However, here we explicate the  controllability of this mechanism demonstrating, for instance, the existence of an upper bound of the height above which the droplet disappears. 
This threshold decreases for larger widths or greater LHY strengths, which is traced back to the appearance of additional bound states in the potential well and the increased droplet localization~\cite{petrov2015quantum} respectively. 
In contrast, a potential barrier enforces the gradual deformation and eventual fragmentation of the droplet into two, a process that occurs faster  for reduced heights in the case of larger widths. 
We also infer the spectral (in)stability of the ensuing droplet configurations in the presence of a potential barrier or well
through Bogoliubov-de Gennes (BdG) linearization analysis. 
The behavior of their collective modes in terms of the central potential defect characteristics is also assessed.   

Upon quenching a potential barrier, we showcase droplet's fragmentation into two counterpropagating segments whose separation and velocity can be tuned by means of the potential height and width. 
A semiclassical model is devised to capture this response, based on energy conservation. 
Meanwhile, the quench excites the droplet fragments which suffer self-evaporation~\cite{Ferioli_evaporation} releasing small density portions. 
On the other hand, the sudden ramp-up of the potential well enforces spatial localization of the droplet which features excitation processes.  
The latter, due to self-evaporation, decay in the course of the evolution letting the droplet to eventually attain a quasi-steady state. 
Monitoring the position variance allows to track the collective motion of the droplet and its excitation frequencies whose values are in-sync with the BdG predictions. 
Finally, insights into the participating microscopic mechanisms and the crucial role of the quantum superposition for the droplet dynamics are unveiled through the single-particle eigenstates of the external potential. 
The dynamical behavior of the eigenstates inherits information about the droplet's excitation processes.

This work is organized as follows. 
Section~\ref{sec:theory} introduces our droplet setting along with its external potential and  eigenstates as well as the reduced single-component eGPE framework. 
Section~\ref{statics} elaborates on the characteristics of the droplet solutions and their stability in the presence of a potential well and barrier.  
In Section~\ref{dynamics1D} we analyze the emergent nonequilibrium droplet dynamics induced by suddenly switching on the potential defect which facilitates the dynamical splitting [Sec.~\ref{dyn_split}] or localization [Sec.~\ref{dyn_loc}] of the original droplet. 
A decomposition of the many-body dynamics in terms of the single-particle eigenstates of the external confinement obtained through exact diagonalization is discussed in Sec.~\ref{SPFs}. 
We conclude and discuss future research extensions based on our results in Section~\ref{conclusions}.

\section{Droplet setting and external potential}\label{sec:theory}

In order to form 1D droplet configurations~\cite{petrov2016ultradilute}, we consider a homonuclear bosonic mixture with masses $m_1=m_2 \equiv m$, intracomponent repulsion strengths  $g_{11}=g_{22} \equiv g>0$ and intercomponent attractive coupling $g_{12}<0$.  
It is known that 1D droplets can be generated under the condition $\delta g= g_{12}+g>0$~\cite{petrov2016ultradilute,mistakidis2023few}. 
Promising candidates for the realization of this mixture are the hyperfine states $\ket{F=1, m_F=-1}$ and $\ket{F=1, m_F=0}$ of $^{39}$K which have been used in relevant three-dimensional experiments~\cite{cabrera2018quantum,semeghini2018self,ferioli2019collisions}. To realize a single-component droplet, we additionally require equal population, i.e. $N_1=N_2\equiv N$, of the involved hyperfine states.

The mixture is confined by a harmonic oscillator with frequency $\omega_x \ll \omega_{\perp}$, where $\omega_x$ ($\omega_{\perp}$) denotes the trapping frequency in the elongated (transverse) direction.
This condition together with $|\mu|/\hbar \ll \omega_{\perp}$ (with $\mu$ denoting the droplet's chemical potential) ensure the 1D character of the system. Namely, due to tight confinement the ensuing excitation energies along the perpendicular $y$, $z$ directions are much larger than the
elongated excitation energies. 
Hence, the atoms are kinematically constrained along the $x$ direction. 
Accordingly, the 1D droplet, in the weakly interacting regime, is adequately described by the eGPE~\cite{petrov2016ultradilute,englezos2025multicomponent} 
\begin{equation}
i\hbar\frac{\partial \Psi}{\partial t} = - \frac{\hbar^2}{2m}\frac{\partial^2 \Psi}{\partial x^2} + \frac{\delta g}{2}|\Psi|^2\Psi - \frac{\sqrt{m}}{\pi \hbar}g^{3/2}|\Psi|\Psi + V(x)\Psi,\label{eGPE}     
\end{equation}
experiencing the competition of repulsive cubic mean-field  interactions and attractive  quantum fluctuations scaling as $\Psi^2 $. 
The latter are responsible for droplet formation~\cite{petrov2016ultradilute,luo2021new}. Here, $\Psi(x,t)$ refers to the 1D droplet wave function and $\hbar$ is the reduced Planck constant. The participating interaction strengths are experimentally tunable through manipulation of the three-dimensional scattering lengths that can be achieved by means of available magnetic fields using Fano-Feshbach resonances~\cite{semeghini2018self,cheiney2018bright} or via confinement induced resonances~\cite{Olshanii_conf_ind,Bergeman_atom_2003} which can be attained by adjusting the transverse confinement.

\begin{figure*}[t!]
\centering
\includegraphics[width=1\linewidth]{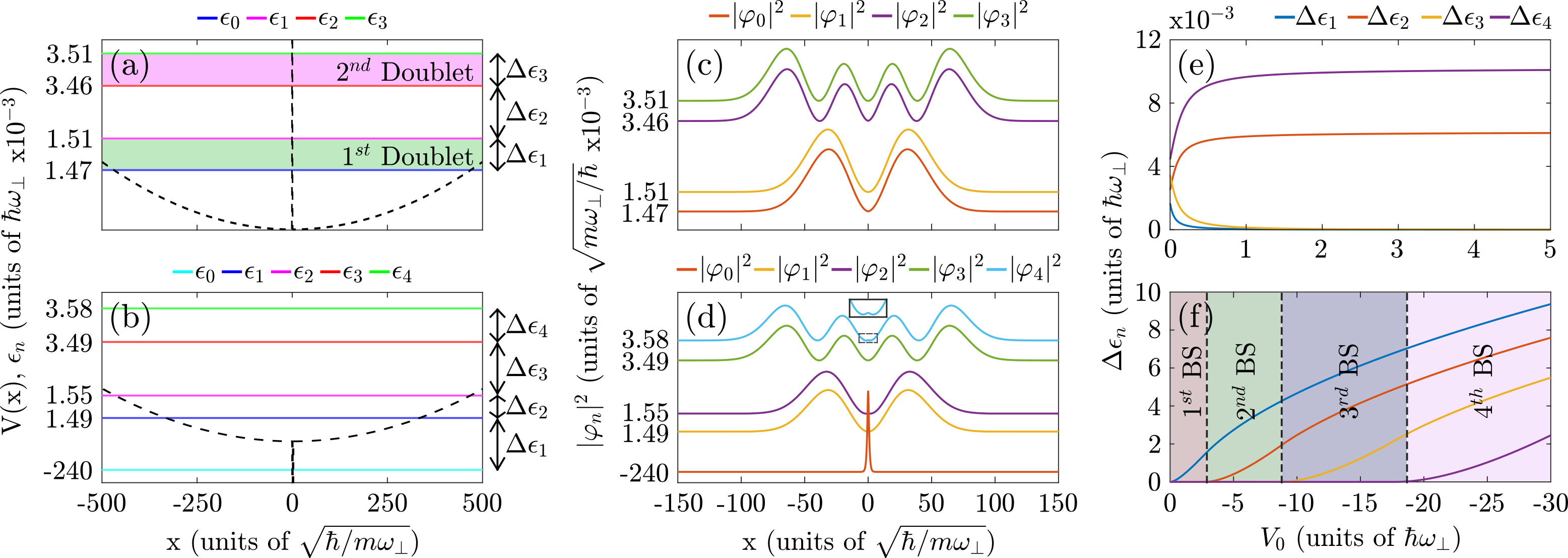}
\caption{Schematic illustration of the external potential described by Eq.~(\ref{potential}) corresponding to (a) a double-well with $V_0>0$ and (b) a harmonic trap with a central dip when $V_0<0$. The single-particle energy levels are marked by $\epsilon_n$,  with $n=0,1,2,\dots$, obtained through exact diagonalization. 
The shaded areas in panel (a) designate the energy doublets. 
The respective eigenstates are  provided in panels (c) [(d)] for $V_0>0$ [$V_0<0$]. 
In both cases, the employed potential characteristics correspond to $V_0=\pm 1$ and $\sigma=0.5$.  
(e), (f) The dependence of the underlying energy gaps among consecutive energy levels, $\Delta \epsilon_n= \epsilon_{n}-\epsilon_{n-1}$, with $n=1,2,\dots$ as a function of the potential depth $V_0$ (see legend). Inset in panel (d) shows a magnification at the center of $\abs{\varphi_4}^2$ to clearly visualize the existent ``kink" of even eigenstates. 
In the double-well, $\Delta \epsilon_n$ among doublets increase for larger $V_0$, while in the potential well case additional bound states (BS) appear for increasing $V_0$ (shaded regions). }
\label{fig:SP_potential}
\end{figure*}

The external potential consists of a loose harmonic trap\footnote{Note that the presence of the harmonic trap is not necessary for the results to be presented below. Similar phenomena occur also with a box confinement in the $x$-direction.} with frequency $\Omega=\omega_x/\omega_{\perp}=0.001$ and a superimposed central defect
\begin{equation}
V(x) = \dfrac{1}{2}m\omega_x^2x^2 + V_0e^{-\frac{x^2}{2\sigma^2}}.   \label{potential}    
\end{equation}
Here, $\sigma$ denotes the spatial width of the central potential defect, and $V_0$ refers to the strength of the potential well  
($V_0<0$) or barrier ($V_0>0$). 
Apparently, the central potential defect  reduces to a delta-like one for $\sigma \to 0$ and $V_0 \to \pm \infty$, while $V_0 \sigma $ is kept fixed.  
Specifically, for $V_0<0$, the overall confinement corresponds to a harmonic trap with a central dip, see Fig.~\ref{fig:SP_potential}(b), facilitating the accumulation of atoms from the droplet at its vicinity since it hosts additional bound states depending on $V_0$ (see also the discussion below). 
This contribution acts against flat-top configurations and its increase may gradually destruct the droplet. 
The potential well hosts a larger number of bound states for larger $V_0$ [Fig.~\ref{fig:SP_potential}(f)], while their energy gap is reduced for smaller $\sigma$. 
On the other hand, for $V_0>0$ a 1D double-well potential [Fig.~\ref{fig:SP_potential}(a)] is realized which favors droplet fragmentation into two segments. 
In this case, a larger $V_0$ results in the increase of the energy gap among consecutive single-particle eigenstate doublets, see also Fig.~\ref{fig:SP_potential}(e).

In what follows, we rescale the time and length scales as well as the droplet wave function with respect to $\omega_{\perp}^{-1}$, $\sqrt{\hbar/m\omega_{\perp}}$ and $(m\omega_{\perp}/\hbar)^{1/4}$ respectively. Similarly, the height, $V_0$, and width, $\sigma$, are expressed in units of $\hbar\omega_{\perp}$, and $\sqrt{\hbar/m\omega_{\perp}}$. 
These yield the dimensionless eGPE  
\begin{equation}
i\frac{\partial \psi}{\partial t} = - \frac{1}{2}\frac{ \partial^2 \psi}{\partial x^2} + \lambda |\psi|^2 \psi - \delta|\psi|\psi + V(x)\psi, 
\label{dimless_eGPE}
\end{equation}
where $\lambda = \frac{\delta g}{2} \sqrt{(\frac{m}{\omega_{\perp}\hbar^3})}$, and $\delta = \frac{1}{\pi} (\frac{mg^2}{\omega_{\perp}\hbar^3})^{3/4}$ represent respectively the strength of the mean-field and the LHY quantum correction. Also, 
$V(x)=\frac{1}{2}\Omega^2x^2+\frac{V_0e^{-x^2/{2\sigma^2}}}{\hbar\omega_{\perp}}$. 
As such, all quantities to be presented below are in dimensionless units. 
Since we aim to examine the non-trivial impact of the potential defect on the droplet, barrier heights satisfying $\abs{V_0} \gg \abs{\mu}$ (with $\mu$ denoting the droplet's chemical potential) are considered. 
Also, the healing length of the droplet, $\xi=\sqrt{\lambda}/\delta=1$, sets a characteristic length scale. 
Accordingly, the central potential widths examined obey $\sigma <\xi$; otherwise droplet destruction occurs already for very small $V_0$ values. 

\begin{figure*}[t!]
\centering
\includegraphics[width=1.0\linewidth]{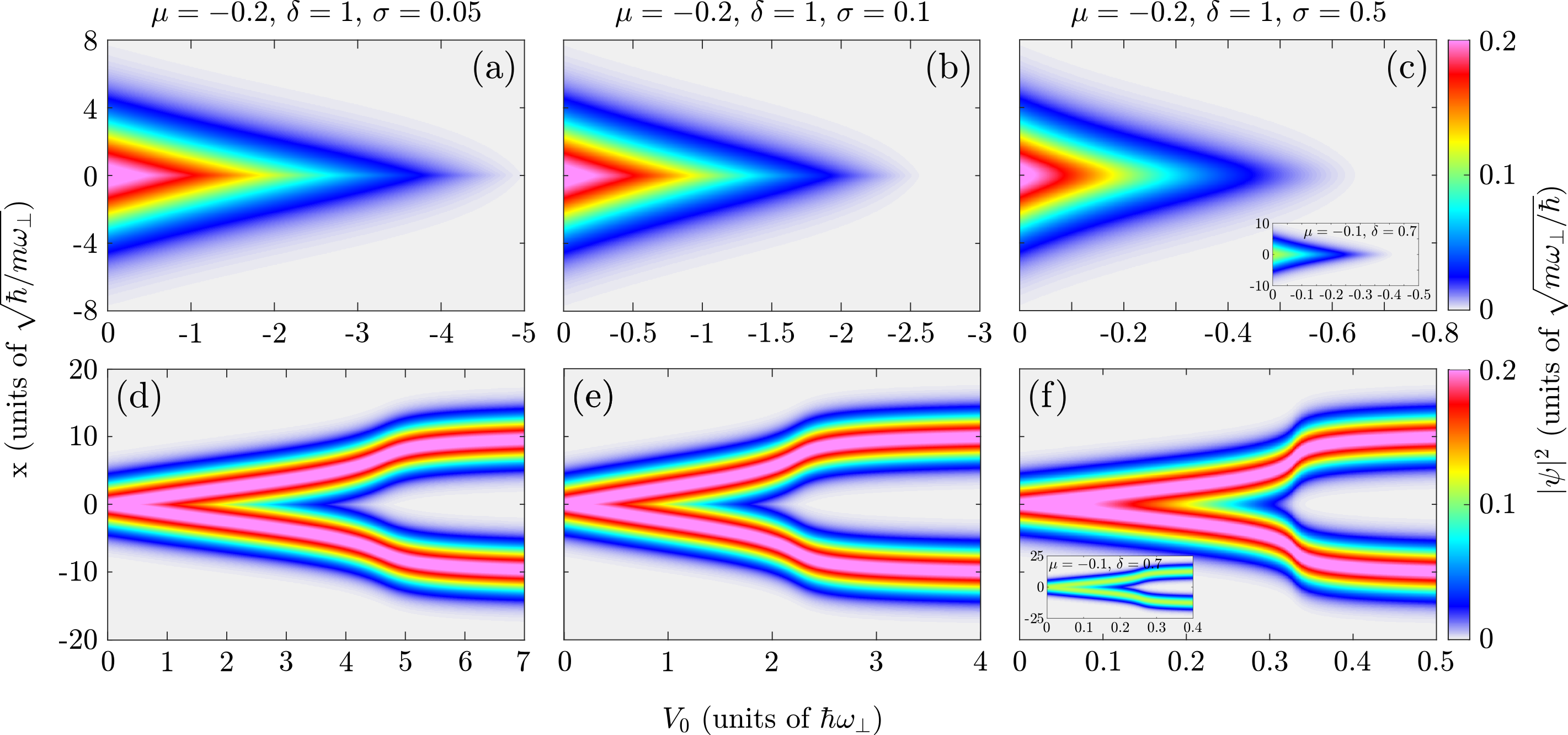}
\caption{Ground-state density configurations of a harmonically trapped droplet in the presence of (a)-(c) a potential well ($V_0<0$) and (d)-(f) a barrier ($V_0>0$) with respect to $V_0$ and different widths $\sigma$ (see legends). 
It is evident that the droplet shrinks in size under the influence of a potential well and eventually disappears, a process that occurs faster for wider potentials. 
A potential barrier leads to two droplet segments forming faster especially for larger widths. 
The chemical potential of the droplet and the strength of the LHY interaction are held fixed (see legends). The insets in panels (c) and (f) show the droplet density distributions when $\mu=-0.1$ and $\delta=0.7$ for the potential well and barrier respectively. A reduced LHY strength facilitates the occurrence of the droplets for larger negative values of $V_0$, while splitting of the structure takes place at smaller positive $V_0$.}
\label{fig:ground_state}
\end{figure*}

Typical trap ratios $\Omega=0.001$ can be achieved through tight transversal confinement $\omega_{\perp}/(2\pi)=1~$kHz which is in a similar range to the 1D experiments of Refs.~\cite{Gorlitz,Moritz,meinert2017bloch,Peregrine_exp} and frequency $\omega_x/(2\pi)=1~$Hz along the elongated axis. The central potential defect is created through a focused laser beam~\cite{Engels_obstacle,Onofrio,lim2022vortex}. 
Accordingly, the range of the potential heights $V_0 \in [0,5]$ and widths $\sigma=0.05$ ($0.5$) considered herein in the employed dimensionless units refer to intensities $\sim [0,240]~{\rm nK}$ and wavelengths $25$ $(254)~{\rm nm}$ respectively. 
Finally, evolution times $t \sim 200$ correspond to $\sim 32~ {\rm ms}$ which are feasible in current droplet experiments~\cite{Cavicchioli,semeghini2018self,d2019observation,cheiney2018bright}. Another interesting realization of our setup  may stem from a highly particle imbalance heteronuclear three-component mixture (not realized in relevant experiments so far). 
Here, the droplet is again composed by two hyperfine states of $^{39}$K and the impurity by a heavy species such as $^{174}$Yb.

To provide evidences about the many-body character of the eGPE droplet dynamics and understand its microscopic characteristics we also perform exact-diagonalization on the level of the single-particle Schr\"odinger equation. 
The respective eigenvalue problem reads $[(-1/2)\partial^2/\partial x^2+ V(x)]\varphi_n(x)=\epsilon_n \varphi_n(x)$, where $\epsilon_n$, $\varphi_n(x)$ represent the eigenvalues and eigenstates of the system. 
This is numerically solved via standard diagonalization\footnote{Using a spatial discretization of  $dx=0.02$, guarantees numerical convergence of the eigenenergies and eigenstates of our setting.} yielding $\epsilon_n$ and $\varphi_n(x)$, with the latter being, of course, orthonormal namely $\braket{\varphi_n|\varphi_m}=\delta_{nm}$. 
For completeness, we remark that the kinetic energy operator is constructed with the finite difference method leading to a tridiagonal matrix representation.

The first few energetically lower single-particle eigenstates of $V(x)$ with the potential barrier and well  are shown in Fig.~\ref{fig:SP_potential}(c), (d). 
It can be seen that in the case of a double well, $V_0>0$, eigenenergy doublets appear with the corresponding  $\varphi_n(x)$ being delocalized over both wells, and hence bearing similarities with Bloch states~\cite{lewenstein2012ultracold}. 
The doublets consist of a symmetric/antisymmetric wave function pair.
As expected~\cite{lewenstein2012ultracold}, the energy gap decreases (increases) with $V_0$ for eigenstates belonging to the same (different) doublet and eventually saturates, see Fig.~\ref{fig:SP_potential}(e).  

On the other hand, for $V_0<0$ there is always a bound state  lying within the potential well  [Fig.~\ref{fig:SP_potential}(d)]. 
This bound state has the form $\varphi_0(x) \sim \sqrt{\kappa}e^{-\kappa \abs{x}}$, with $\kappa = \abs{V_0}\sigma \sqrt{2\pi}$ which in the limiting case of ($\sigma \to 0$, $V_0\to \infty$) converges to the relevant bound state in a delta potential.
Higher-excited states are delocalized, with their extent being dictated by the harmonic oscillator length, $l_x= \Omega^{-1/2} = 31.6$.
The odd $\varphi_n$ eigenstates ($n=1,3,\ldots$) are not affected by the central dip since they possess a node at this location.
This is manifested by their eigenenergies being close to the values in the absence of a dip, $(n+1/2)10^{-3}\hbar \omega_{\perp}$ [Fig.~\ref{fig:SP_potential}(d)].
In contrast, the even eigenstates ($n=2,4,\ldots$) are affected by the potential well, exhibiting a suppressed peak at the center [inset of Fig.~\ref{fig:SP_potential}(d)].
Moreover, their eigenenergies are shifted to lower values compared to the case with $V_0=0$.
The energy gaps among successive eigenstates increase for larger $V_0$ and additional bound states appear as can be inferred from Fig.~\ref{fig:SP_potential}(f).

To obtain the stationary states of the many-body setting analyzed below, we compute the time-independent version of Eq.~(\ref{dimless_eGPE}) utilizing a fixed point iterative Newton scheme~\cite{kelley2003solving} in the presence of von Neumann boundary conditions. 
Additionally, in order to monitor the nonequilibrium quantum dynamics of the droplet we solve Eq.~(\ref{dimless_eGPE}) in real-time through a fourth-order Runge-Kutta integrator exploiting a second-order finite differences method for the spatial derivatives. Typical values of the used spatial and time discretizations are $dx=0.005$ and $dt=10^{-6}$ respectively.

\section{Static droplet configurations}\label{statics}

First, we start by investigating the impact of the potential well of Eq.~(\ref{potential}) on the harmonically trapped 1D droplet. 
Throughout this work, a weak harmonic trap of frequency $\Omega=0.001$ is used which naturally imposes an additional length scale, i.e. the harmonic oscillator length $l_x=\Omega^{-1/2}$ atop the free space droplet structure. This acts  against the formation of spatially extended flat-top states~\cite{englezos2023correlated} but also enforces an overall shift of the droplet's binding energy towards less negative values~\cite{katsimiga2023solitary}. 
Our aim, here, is to explore structural deformations, regions of existence and stability properties of the emergent droplets experiencing an additional central potential well or barrier  which may emulate a heavy impurity atom inside the droplet. 
Below, our analysis for the stationary states mainly  pertains to specific values of the chemical potential, $\mu=-0.2$, strength of the LHY attraction, $\delta=1$ and average mean-field repulsion $\lambda=1$.  
The fact that $\mu<0$ reflects the bound state character of the many-body state, while our focus is placed on $\abs{V_0} \gg \abs{\mu}$ such that the potential defect significantly influences the droplet configuration. 
However, we have also verified the generalization of our results upon appropriate variations of these parameters (e.g. $\mu=-0.05,-0.1,-0.2222$, $\delta=0.7$) and suitable comments are provided within our description.

\subsection{Density distribution of the droplets} 

To visualize the structural modifications of the harmonically trapped 1D droplet due to the presence of the potential well we inspect the density, $n(x)=\abs{\psi(x)}^2$, of the system for different barrier characteristics, see Fig.~\ref{fig:ground_state}. 
As discussed in Sec.~\ref{sec:theory} and depicted in Fig.~\ref{fig:SP_potential}(f) the potential well accommodates additional bound states whose number increases for larger $V_0<0$. 
Hence, atoms from the droplet accumulate in these bound states substantially distorting its spatial distribution. 
Prototype examples confirming this behavior are shown in Figs.~\ref{fig:ground_state}(a)-(c) for $\mu=-0.2$ and distinct values of $\sigma$ in terms of $V_0$.  
Specifically, the original (at $V_0=0$) flat-top droplet gradually shrinks in size and its peak density decreases for more negative values of $V_0$ since a more appreciable amount of atoms are trapped within the central potential well. 
In fact, the computed structurally deformed droplet configurations follow the analytical solution reported in Refs.~\cite{debnath2023interaction,abdullaev2020bosonic} in the presence of a delta potential  
\begin{equation}
\Psi_{{\rm D1}}(x,\mu)=\frac{3 \mu}{1+\sqrt{1+9 \mu/2} \cosh[\sqrt{-2 \mu}(\abs{x}+ \xi_D)]}, \label{droplet_delta}    
\end{equation}
where $\xi_D$ is the amended droplet's healing length. We have confirmed this excellent agreement (not shown) upon fitting Eq.~(\ref{droplet_delta}) to the numerically obtained non-zero eGPE waveforms presented in Fig.~\ref{fig:ground_state}(a)-(c). 
The above-mentioned trapping property depends crucially on the height of the central potential well as the number of bound states at the single-particle level increases for larger $V_0$ and fixed $\sigma$, $\mu$, see also Fig.~\ref{fig:SP_potential}(f). 
Hence, the effect of the central potential well is more prominent for larger $\sigma$, compare in particular Figs.~\ref{fig:ground_state}(a), (b) and (c), leading eventually to the termination of the droplet solution. 
For instance, in the case of $\mu=-0.2$, $\delta=1$, displayed in Fig.~\ref{fig:ground_state}(a)-(c), the droplet disappears at $\abs{V_0}=5,~2.7,~0.68$ when $\sigma=0.05,~0.1,~0.5$ respectively. 
The above thresholds match closely the ones derived for a droplet in a Dirac delta potential, which exists only for $ -\mu > \abs{V_0}^2 \sigma^2 \pi$~\cite{debnath2023interaction}.

A similar response takes also place for different values of the chemical potential (not shown) and the LHY strength as long as the droplet solution exists. 
Recall that the latter occurs only for $\abs{\mu} > 2\delta^2 / 9$~\cite{katsimiga2023solitary} and thus for $\delta = 0.7$ for instance, droplets exist at $\mu = -0.1$. 
For these parameters, the impact of the potential well on the droplet [inset of Fig.~\ref{fig:ground_state}(c)] is less drastic compared to $\delta=1$, leading to the destruction of the structure for relatively larger values of $V_0$. 
For example, fixing $\mu=-0.1$ we find that the solution terminates at $V_0\approx -0.42$ [$V_0\approx - 1.78$] for $\delta=0.7$ and at $V_0=-0.40$ [$V_0=-1.74$] for $\delta=1$ when $\sigma=0.05 \ll \xi$ [$\sigma=0.1 < \xi$]. 
This is because the droplet distribution for a specific $\mu$ becomes broader and less peaked as $\delta$ decreases due to the smaller 1D LHY attraction (with the mean-field repulsion dominating). 
Thus, it requires a larger negative $V_0$ to achieve the same number of atoms in the bound states for broader droplet distributions. 

On the other hand, the free droplet solution in the absence of the central potential has been demonstrated to exist for $-0.222<\mu<0$~\cite{astrakharchik2018dynamics,katsimiga2023solitary} when $\delta =1$. 
In this case, keeping all parameters fixed and tuning $\mu$ towards the lower bound a transition from Gaussian type to flat-top droplet configurations occurs. 
As such droplets characterized by less negative $\mu$ are more susceptible to the presence of the additional potential defect and cease to exist for smaller negative values of $\mu$ since they accommodate a relatively smaller amount of atoms as was shown, for instance, in Refs.~\cite{astrakharchik2018dynamics,katsimiga2023solitary}.    

\begin{figure*}[t!]
\centering
\includegraphics[width=1\linewidth]{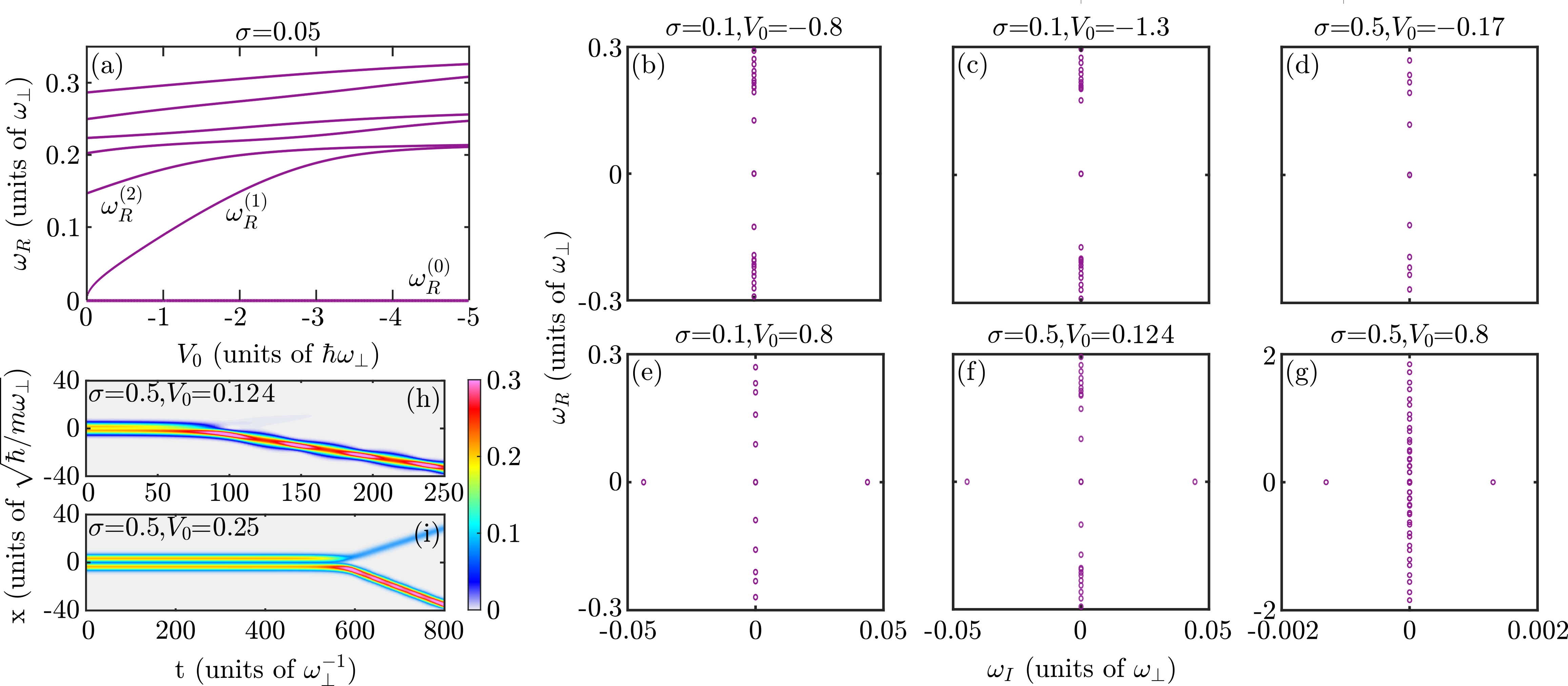}
\caption{(a) Real part of the BdG excitation spectra for the harmonically trapped droplet under the influence of an external potential well as a function of its height, $V_0$. The behavior of the collective excitation branches, labeled $\omega^{(n)}_R$ with $n=0,1,2,\ldots$, is shown. 
Selected imaginary versus real parts of the ensuing  BdG droplet spectra in the case of (b)-(d) potential well and (e)-(g) a potential barrier for different potential characteristics (see legends). The presence (absence) of finite imaginary parts in the case of a potential barrier (well) reflects the unstable (stable) nature of the droplet configurations. 
(h), (i) Dynamical evolution of the perturbed stationary configurations at selected $V_0$ values of the potential barrier.
In all cases, the chemical potential, $\mu=-0.2$, and the LHY strength is $\delta=1$ (see legends).
}
\label{fig:spectra}
\end{figure*}

Turning to central potential barriers, characterized by $V_0>0$, we observe a completely different effect on the droplet distribution. 
Ramping-up the barrier progressively enforces a splitting of the single droplet. 
Initially, this process yields a shallow two-hump configuration [see, e.g., Fig.~\ref{fig:ground_state}(d) for $0.5<V_0<4$] leading eventually to a well-separated two droplet distribution [see, for instance, Fig.~\ref{fig:ground_state}(d) for $V_0>5$].  
As can be seen from the density distributions in Figs.~\ref{fig:ground_state}(d)-(f), the aforementioned splitting mechanism occurs naturally earlier in terms of $V_0$ for larger widths because the exerted force from the barrier to the droplet is increased.

To infer the droplet nature of the density distribution at large barrier heights, $V_0>5$ we also perform a fitting of each droplet fragment to the analytically known~\cite{astrakharchik2018dynamics,tylutki2020collective,katsimiga2023interactions} 1D droplet solution in the absence of an external potential (i.e. $V(x)=0$ in Eq.~(\ref{dimless_eGPE}))
\begin{equation}
\Psi_{{\rm D2}}(x,\mu)= \frac{3\mu}{\delta \left(1+\sqrt{1+\frac{9\mu}{2\delta^2}}\cosh\left(\sqrt{-2\mu} (x - x_0)\right)\right)}. \label{droplet_solution}
\end{equation}
Here, $\mu$ denotes the background chemical potential, $x_0$ is the center of each droplet segment and $\delta$ refers to the LHY strength. 
These parameters are determined by the fitting to the eGPE waveform.   
An excellent agreement between our numerically obtained two-droplet solution and the analytical estimate can be extracted (not shown for brevity), thus confirming the droplet character of the former. 
A careful inspection of Figs.~\ref{fig:ground_state}(d)-(f) also reveals that the participating density fragments in the two-droplet configurations are thicker for reduced potential barrier width, compare for instance Figs.~\ref{fig:ground_state}(d) and (f). 
This is expected since in the case of a ``thicker" barrier the local harmonic oscillators possess tighter frequencies. 
As such, the trapping becomes more prominent resulting in relatively shrinked droplets. 

On another note, a reduced LHY strength [inset of Fig.~\ref{fig:ground_state}(f)] produces wider droplets which split earlier compared to larger $\delta$ values. 
This is attributed to the smaller LHY attraction which favors droplets of somewhat larger width and smaller peak density.

\subsection{Stability analysis of the droplet structures}

Having analyzed the deformations of the droplet configurations due to the presence of the central potential defect and their parametric regions of existence, we next examine the spectral stability of the ensuing solutions through a BdG  analysis~\cite{stoof2009bose}. 
Note here that the excitation spectrum of symmetric droplets has been previously discussed in 1D~\cite{tylutki2020collective,katsimiga2023interactions} and 2D~\cite{Fei_2Dspec,Bougas_vortex_drops} but also for genuine two-component droplets~\cite{Charalampidis_2comp_drops} in both free space or under the influence of a harmonic trap, see also Refs.~\cite{katsimiga2023solitary,liu2022vortex} for estimating the impact of nonlinear excitations. 
On the other hand, the stability of droplets experiencing a central potential defect atop the harmonic trap remains unprecedented to the best of our knowledge. 

To linearize the eGPE~(\ref{dimless_eGPE}), we deploy a small amplitude ($\epsilon \ll1$) perturbation to the stationary solutions, $\psi^{(0)}(x)$, according to 
\begin{equation}
\tilde{\psi}(x)=e^{-i \mu t}
\left[\psi^{(0)}+\varepsilon\left(ae^{i\omega t}+b^{*}e^{-i\omega^{*}t}\right)\right]. 
\label{BdG_ansatz}
\end{equation}
In this expression, [$a(x)$, $b(x)$] represents the eigenvector with eigenfrequency $\omega=\omega_{R}+i\,\omega_{I}$. Here, $\omega_R$ ($\omega_I$) corresponds to the real (imaginary) part of the  eigenfrequency.
Inserting the aforementioned ansatz into Eq.~(\ref{dimless_eGPE}) and keeping terms of $\mathcal{O}(\varepsilon)$ allows the  extraction of a linearized set of equations  that has been reported, for instance, in Refs.~\cite{tylutki2020collective,katsimiga2023solitary,katsimiga2023interactions}. 
The relevant equations are subsequently solved numerically yielding the ensuing discrete eigenfrequencies, due to the presence of the external trap, and eigenvectors providing access to the spectrum of the system and thus allowing us to deduce its stability properties. 
It is apparent that a configuration is deemed stable as long as $\omega_I=0$; otherwise instabilities occur. 
On the other hand, $\omega_R$ captures the collective excitation frequencies of the respective configuration and their behavior as a function of a system parameter.

The real part of the BdG spectrum, 
$\omega_R$, is illustrated in Fig.~\ref{fig:spectra}(a) for droplet distributions with $\mu=-0.2$ and $\delta=1$ upon parametric variation of the height of the potential well characterized by a width $\sigma=0.05 \ll \xi$. 
The dependence of the first few eigenfrequencies, $\omega^{(n)}_R$ with $n=0,1,\ldots, 6$, on the potential depth (or height) can be discerned. 
Specifically, the trajectory traced by 
$\omega^{(0)}_R=0$ corresponds to the 
zero-eigenfrequency mode that is related to an infinitesimal phase shift of the droplet's wave function (phase invariance).
The consecutive one, i.e. $\omega^{(1)}_R$,
is associated with the droplet's center-of-mass displacement (dipole mode~\cite{tylutki2020collective,englezos2023correlated}) and the trajectory traced by $\omega^{(2)}_R$ corresponds to the collective breathing mode~\cite{tylutki2020collective,astrakharchik2018dynamics} of the droplet solutions. 
For increasing $V_0$ while keeping $\sigma$ fixed, it turns out that $\omega^{(n)}_R$ are shifted to larger values. For example, 
$\omega^{(2)}_R \approx 0.181$ for $V_0=-1$ 
becoming $\omega^{(2)}_R \approx 0.214$ for $V_0=-4.99$ [Fig.\ref{fig:spectra} (a)] at the termination point of the droplet solution.
The same holds true upon fixing $V_0$ e.g. to 
$V_0=-0.2$ and varying $\sigma$. 
In this case, we find that 
$\omega^{(2)}_R \approx 0.155$ for $\sigma=0.05$, acquiring the value of $\omega^{(2)}_R \approx 0.197$ for $\sigma=0.5$. 

Importantly, all solutions identified herein stemming from a potential well are stable irrespective of its width and height. 
This is verified by the zero imaginary part of the eigenfrequencies. 
To explicate this general property we present selective eigenspectra in the $\omega_R-\omega_I$ plane for different potential well heights and widths in Figs.~\ref{fig:spectra}(b)-(d). 
It can be readily seen that in all cases $\omega_I=0$ (being typically of the order of $\approx  10^{-8}$ herein) which supports the spectral stability of these deformed droplet solutions. 
To independently verify the stability 
predicted by the BdG analysis, we confirmed that
the solutions remain intact when freely evolved up to $t= 10^4$ (in dimensionless units).

Turning to the potential barrier case, a completely different deformation of the droplet configuration occurs. 
Recall that as we depart from $V_0=0$, namely from the harmonically trapped state, the original droplet gets depleted, with a gradually increasing density dip being imprinted around its core due to the finite and progressively higher value of $V_0$.
This depletion leads to the segregation of the stationary configuration into two fragments whose separation increases for larger $V_0$ [see Fig.\ref{fig:ground_state}(d)-(f)]. 
This deformed configuration, is found to be unstable for all considered non-negligible  values of $V_0$ [see e.g. Fig.~\ref{fig:spectra}(e)-(g)] which alter the original droplet structure. 
The maximum growth rate of the ensuing instability corresponds to $\omega_I=0.044$ occurring at $V_0=0.124$ e.g. for $\sigma=0.5$.
The relevant real vs imaginary BdG spectrum is presented in Fig. \ref{fig:spectra}(f) where the imaginary eigenfrequency pair can be seen.
However, as segregation progresses, the relevant eigenfrequencies acquire $\omega_I\sim \mathcal{O}(10^{-4})$ [Fig. \ref{fig:spectra}(g)] asymptotically tending to two fairly separated stable individual droplets (see also the discussion below). 

Among the modes that appear in the relevant real part of the BdG spectrum as a function of $V_0$ (not shown for brevity) it is $\omega^{(1)}_R$ that is associated, via its eigenvector, with a symmetry breaking of the deformed droplet-like structure that leads to particle imbalanced entities. 
To showcase the aforementioned stability analysis findings, in Fig.~\ref{fig:spectra}(h), (i) we monitor the dynamical evolution of a perturbed configuration. 
The perturbation consists of adding the relevant eigenvector to the stationary solution.
Dynamical destabilization of the initial depleted configuration, when perturbed with the eigenvector corresponding to the maximal growth rate, manifests itself at early times through a dramatic particle redistribution. This leads to a single excited droplet-like entity moving towards the trap edge [Fig.~\ref{fig:spectra}(h)]. 
However, for fairly separated fragments, instability leads into two counterpropagating particle imbalanced droplet-like waveforms [Fig.~\ref{fig:spectra}(i)]. 
Similar mass transfer mechanisms among two colliding 1D droplets have been seen in Refs.~\cite{astrakharchik2018dynamics,katsimiga2023interactions}. 
\begin{figure*}[t!]
\centering
\includegraphics[width=1\linewidth]{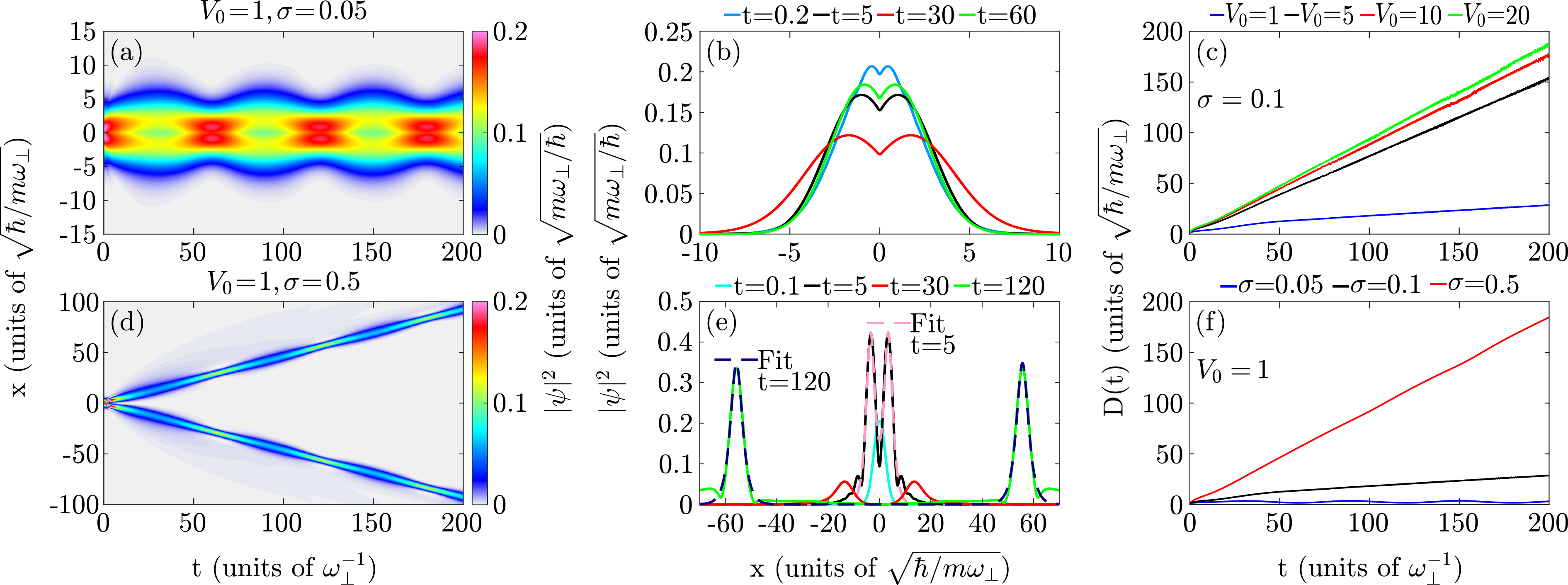}
\caption{Density evolution of the droplet after ramping-up the central potential barrier of height $V_0=1$ and width (a) $\sigma=0.05$, (d) $\sigma=0.5$. (b) [(e)] Selected density profiles taken from (a) [(d)] at specific time-instants (see legends), along with the droplet fits (dashes lines) obtained through Eq.~(\ref{droplet_solution}). 
The quench tends to produce two counterpropagating droplet fragments. 
The latter move further apart and their separation $D(t)$ increases both by (c) varying $V_0$ and fixing the width (see legend), and (f) varying $\sigma$ while $V_0=1$.
All other parameters are the same as in Fig.~\ref{fig:spectra}.}
\label{fig:dynamics_repulsive}
\end{figure*}

\section{Droplet dynamics}\label{dynamics1D}

The analysis of the existence and stability properties of static droplet configurations under the influence of a central potential barrier or well paves the way to probe specific dynamical features of the original harmonically trapped droplet. 
This investigation is certainly relevant for guiding  possible experimental endeavors aiming to explore the effect of the potential defect on the droplet distributions. 
For instance, according to the above-discussed phenomenology it is possible to utilize the potential barrier to observe droplet fragmentation or the potential well to excite internal modes of the droplet. 
Also, the BdG excitation spectrum can be used to identify the nature of the dynamically excited modes. 
Below, we showcase such dynamical traits with the aid of a quench of the potential defect on the initial harmonically trapped droplet distribution.

\subsection{Dynamical splitting of the droplet}\label{dyn_split} 

First, we examine the quench-induced dynamics caused by a sudden ramp-up of the potential barrier  characterized by specific widths, $\sigma$, and heights, $V_0$. 
To be concrete, we consider an initially harmonically trapped droplet with $\mu=-0.2$ (i.e. close to the lower $\mu$ limit of existence of  
droplet solutions) experiencing an external potential described by Eq.~(\ref{potential}) with $V_0=0$ and $\Omega=0.001$. 
Subsequently, we quench both $V_0>0$ and $\sigma$ to realize a double-well potential as shown in Fig.~\ref{fig:SP_potential}(a). 
Revisiting the corresponding stationary configurations illustrated in Fig.~\ref{fig:ground_state}(d)-(f), it is anticipated that such a quench facilitates, in general, the splitting of the droplet into two segments. 
In what follows, we initially discuss quenches of the same potential height, $V_0=1$, and different widths $\sigma$ providing an overview of the emergent dynamical features. Subsequently, generalizations of our arguments are also offered for varying height of the potential barrier.

To appreciate the impact of the potential width on the dynamical response, we present the time-evolution of the droplet's density for fixed $V_0=1$ and different postquench values of $\sigma$ in Figs.~\ref{fig:dynamics_repulsive}(a), (d). 
It can be observed that a relatively small width, e.g. $\sigma=0.05 \ll \xi$ shown in Fig.~\ref{fig:dynamics_repulsive}(a), is not sufficient to split the original droplet. 
This is anticipated from the relevant 
ground-state distributions where for $\sigma=0.05$ the droplet fragments after $V_0>4.5$.  
However, the ramping of the potential barrier suffices to significantly perturb the droplet distribution by imprinting a permanent density dip at the droplet core, see also the relevant density profiles at different time instants in Fig.~\ref{fig:dynamics_repulsive}(b), since atoms are repelled from this region. 
Notice here the resemblance between the effect of the potential barrier on the atom distribution with the impact of repulsive impurities immersed in a 1D Bose gas~\cite{grusdt2024impurities,mistakidis2023few}.   
Additionally, the presence of the potential well triggers a collective motion of the droplet. 
This practically refers to a breathing mode of the droplet whose distribution periodically expands and contracts [see also Fig.~\ref{fig:dynamics_repulsive}(b)] possessing a frequency $\omega \approx 0.125 \approx \omega_R^{(2)}$. 
We have checked that this value matches the prediction of the suitable branch appearing in the corresponding real part of the appropriate BdG spectrum (not shown).

On the other hand, increasing the potential barrier's width results
in a dramatically different dynamical response, see Fig.~\ref{fig:dynamics_repulsive}(d).  
Indeed, larger widths facilitate droplet separation due to the increased force exerted by the potential, in line with our predictions at the stationary level [see Fig.~\ref{fig:ground_state}(f)]. 
Therefore, the original harmonically trapped droplet gradually splits into two counterpropagating fragments traveling towards the trap edges as shown in Figs.~\ref{fig:dynamics_repulsive}(d), (e). 
These are droplet fragments, as evidenced
by the fitting of their waveforms to the analytical droplet solution [Eq.~(\ref{droplet_solution})] shown in Fig.~\ref{fig:dynamics_repulsive}(e) at $t=120$. Simultaneously, each of the nucleated droplet fragments becomes excited as can be inferred from its vibrational pattern over time in Fig.~\ref{fig:dynamics_repulsive}(d). 
This excitation dynamics of the droplet fragments is also accompanied by the emission of relatively small density portions, see in particular the blue faint density ripples surrounding the droplet segments in the course of the evolution. 
This phenomenon is a manifestation of the droplet's self evaporation~\cite{Ferioli_evaporation} due to its excited nature. 

\begin{figure*}[t!]
\centering
\includegraphics[width=1\linewidth]{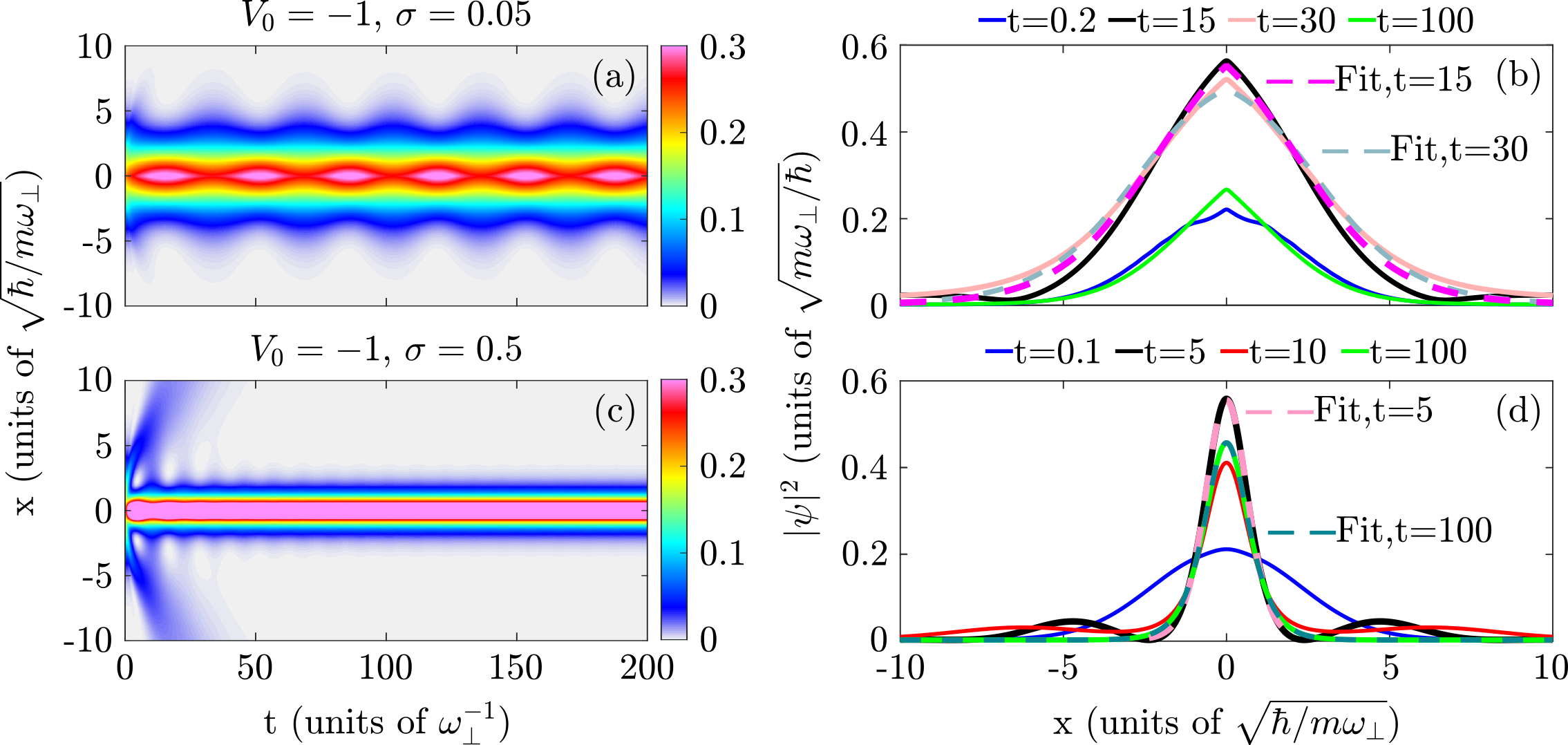}
\caption{Dynamics of the droplet density following a quench of the central potential well from height $V_0=0$ to $V_0=-1$ for various widths (a) $\sigma=0.05$ and (c) $\sigma=0.5$. Relative small amplitude quenches trigger a collective motion of the droplet [panel (a)]. 
However, larger amplitudes lead to self-evaporation along with droplet excitation at short times which eventually decays tending to a quasi-steady state at long times [panel (c)]. (b) [(d)] Instantaneous densities taken from (a) [(c)]. Panels (b), (d) demonstrate also fits of the analytical droplet solution described by Eq.~(\ref{droplet_delta}) and Eq.~(\ref{droplet_solution}) respectively, to the eGPE obtained wave functions at specific time-instants (see legend). Adequate agreement occurs, excluding the tails of the distributions. The chemical potential used is $\mu=-0,2$ and the strength of the LHY $\delta=1$.}
\label{fig:dynamics_attractive_shallow}
\end{figure*}

In order to track the droplet trajectory or the distance among the nucleated droplets depending on the  characteristics of the quenched potential barrier we compute 
\begin{equation}
D(t)= 2 \int _0^{L/2} dx~x~\abs{\psi(x,t)}^2.\label{position} 
\end{equation} 
Here, $L$ denotes the size of our computational domain and the factor of two takes into account the fact that the counterpropagating droplet fragments are symmetric with respect to $x=0$, implying that $D(t)$ measures their distance in time. 
This observable can be experimentally monitored via \textit{in-situ} measurements~\cite{Ronzheimer}. 
The time-evolution of $D(t)$ is shown in Fig.~\ref{fig:dynamics_repulsive}(c), (f) for various postquench potential barrier amplitudes, $V_0$, and widths, $\sigma$. 
It can be easily deduced that for quenches of fixed width, e.g. $\sigma=0.5 \sim \xi$, but different potential barrier height the distance among the generated droplets increases over time in an almost linear manner, see Fig.~\ref{fig:dynamics_repulsive}(c). 
Particularly, larger amplitude quenches yield in general a larger inter-droplet separation accompanied by an increased velocity of the nucleated droplets. 
This is related to the fact that barriers with larger heights exert a stronger force on the original droplet resulting in more violent separation.
The relatively small amplitude oscillations imprinted on $D(t)$ essentially reflect the excited nature of the nucleated droplets, see also Fig.~\ref{fig:dynamics_repulsive}(d). 
A similar response takes place by assuming fixed height of the potential barrier and variable width [see Fig.~\ref{fig:dynamics_repulsive}(f)]; increasing $\sigma$ leads to larger separations and velocities of the generated droplets. 
This is again traced back to the increased force from the barrier to the droplet. 

An energy argument can be employed to intuitively understand the inter-droplet separation, namely that the potential energy after the barrier quench is converted to kinetic for the two fragments.
The potential energy, $E_p = \int dx~ V(x) $ $ \abs{\psi(x,0)}^2 $, can be approximated as $E_p \simeq \abs{\psi(0,0)}^2 V_0 \sigma \sqrt{2\pi}$, where we assumed that the initial droplet is constant within $\abs{x}<\sigma$.
Subsequently, $E_p$ is equally partitioned into the two droplet fragments, each acquiring a velocity $v \simeq \sqrt{E_p}$.
This implies that $D(t)=2vt$ is a good approximation only for large $\sigma$ and moderate $V_0$. For the other cases, the droplet fragments become significantly excited, and thus the above energy argument becomes invalid. 
Summarizing, we can infer that tuning either the height or the width of the potential barrier provides a knob to control the inter-droplet distance and each fragment's velocity.

\begin{figure*}[t!]
\centering
\includegraphics[width=1\linewidth]{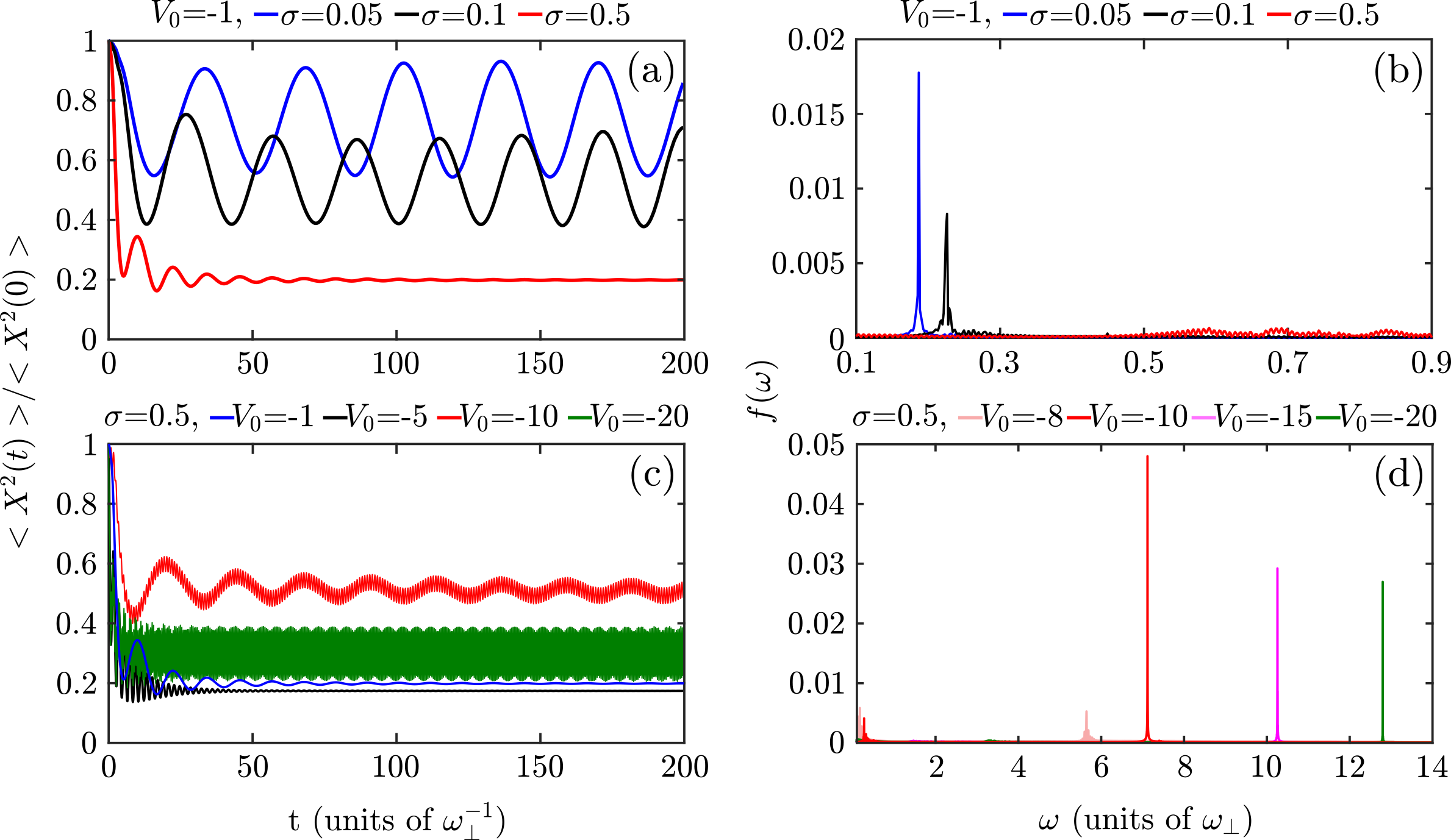}
\caption{Time-evolution of the quenched droplet position variance, $\braket{X^2(t)}/\braket{X^2(0)}$, for (a) fixed height, $V_0=-1$ and various widths (see legend) as well as (c) a fixed width, $\sigma=0.5$, and different heights (see legend) of the potential well. The recurring behavior of the variance signals the collective mode excitation of the droplet with decaying amplitude. For relatively large quenches, the system tends to a quasi-steady state at long evolution times captured by the almost constant behavior of the  variance for $t>50$. The frequency spectrum of the position variance is depicted in panel (b) [(d)] for distinct widths [heights] (see legends). The well-defined peaks in the spectrum refer to the collective motion of the droplet in the modified harmonic trap. For all settings, $\mu=-0.2$ and $\delta=1$.}
\label{fig:dynamics_variance}
\end{figure*}

\subsection{Droplet localization and self-evaporation}\label{dyn_loc}

Next, we proceed by considering quenches of the potential well atop the initially harmonically trapped droplet. 
This protocol is expected to induce dynamical localization of the droplet structure since it favors the trapping of a certain number of atoms into the bound state created by the central well. 
Such a behavior is, of course, further supported by the ground state analysis
of the system illustrated in Fig.~\ref{fig:ground_state}(a)-(c). 

\begin{figure*}[t!]
\centering
\includegraphics[width=1\linewidth]{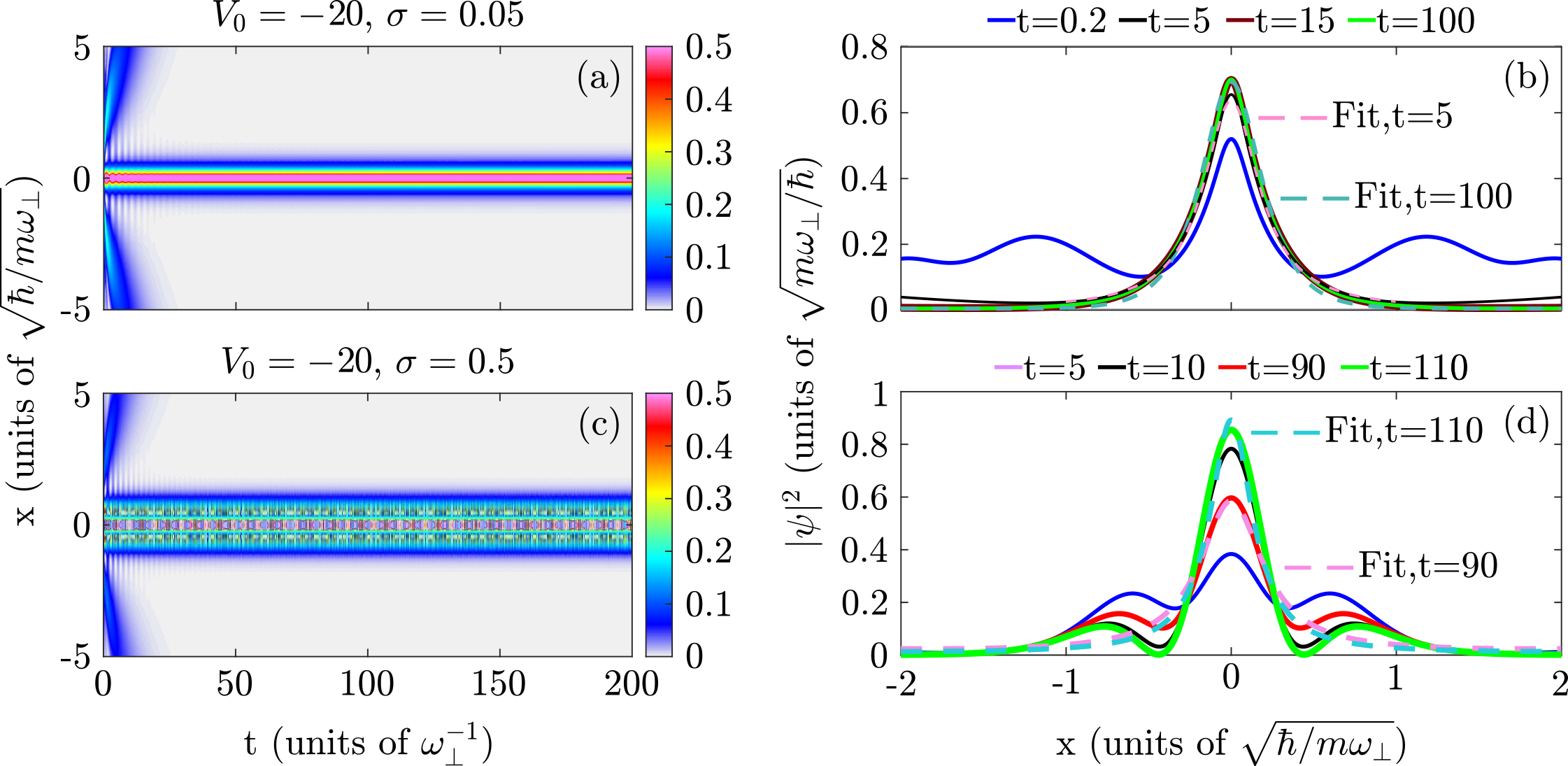}
\caption{Droplet density dynamics using a quench of the central potential well with height, $V_0=-20$, from width $\sigma=0$ to (a) $\sigma=0.05$ and (c) $\sigma=0.5$. Self-evaporation is observed in both cases with the small width quench resulting in a quasi-steady state and the large width one to a highly excited droplet. 
(b) [(d)] Density profiles of (a) [(c)] at selected time-instants (see legends). Panels (b), (d) showcase fits of the analytical droplet solution [Eq.~(\ref{droplet_solution})] to the eGPE distribution at distinct evolution times on the density level (see legend). Good  agreement is observed at the droplet core. Here, the droplet parameters are the same as in Fig.~\ref{fig:dynamics_variance}.}
\label{fig:dynamics_attractive_strong}
\end{figure*}

Characteristic examples of the emergent droplet density  dynamics after a quench to fixed $V_0=-1$ and distinct potential well widths are demonstrated in Fig.~\ref{fig:dynamics_attractive_shallow}(a), (c). 
Focusing on relatively small widths, such as $\sigma=0.05 \ll \xi$, we observe a localization tendency of the perturbed droplet where a significant portion of atoms resides in the vicinity of the potential well, see Fig.~\ref{fig:dynamics_attractive_shallow}(a), performing a breathing motion.   
The aforementioned localization trend of the droplet can be clearly discerned in the instantaneous density profiles, displayed in Fig.~\ref{fig:dynamics_attractive_shallow}(b),  exhibiting an overall spatially shrinked distribution. 
The latter clearly interchanges, during the evolution, between a configuration having a spatially distorted peak structure [see e.g. the profile at $t=0.2$ in panel (b)] around $x=0$ where the potential well lies, and a ``triangle" shaped density resembling a peakon [see, for instance, panel (b) at $t=100$]. 
The latter structures\footnote{They are also reminiscent of the bound state solution of a single-particle in an attractive delta potential ($\psi(x) \sim \sqrt{\kappa} e^{-\kappa  \abs{x}}$, $\kappa = \abs{V_0} \sigma \sqrt{2\pi})$, see also the discussion in Sec.~\ref{sec:theory}.} are similar to the analytical droplet solution in the presence of a delta potential as can be inferred from the fits provided in  Fig.~\ref{fig:dynamics_attractive_shallow}(b). 
Moreover, the above-discussed response is reminiscent of the reaction of a bosonic host gas attractively interacting with impurity atoms (especially of infinite mass), giving rise to attractive polaron states~\cite{grusdt2024impurities,mistakidis2020many}.

The accumulation of atoms towards the minimum of the potential well causes also a collective motion of the droplet cloud within the deformed trap corresponding to a breathing motion with frequency $\omega \approx 0.18 $.  
The latter can be estimated by exploiting the position variance of the bosonic cloud which is given by 
\begin{equation}
\braket{X^2(t)}= \braket{\psi(x,t)|\hat{x}^2|\psi(x,t)}.\label{variance} 
\end{equation}
In this expression, we assume that $\braket{\psi(x,t)|\hat{x}|\psi(x,t)}=0$ since quenching the potential well preserves the parity symmetry of the droplet cloud with respect to $x=0$, a condition that has been also numerically confirmed. 
Note also that this observable can be experimentally probed through \textit{in-situ} absorption measurements~\cite{Ronzheimer}. 
The time-evolution of $\braket{X^2(t)}/\braket{X^2(0)}$ for the particular quench is presented in Fig.~\ref{fig:dynamics_variance}(a). 
It clearly captures the periodic expansion and contraction dynamics of the droplet cloud. 
Moreover, its frequency spectrum depicted in Fig.~\ref{fig:dynamics_variance}(b), $f(\omega) = \int dt ~ e^{-i \omega t} \braket{X^2(t)}/\braket{X^2(0)}$, shows a clear peak at {\bf $\omega \approx 0.18 \approx \omega^{(2)}_R$} referring to the breathing mode and being in line with the BdG predictions, see $\omega^{(2)}_R$ for $V_0=-1$ in Fig.~\ref{fig:spectra}(a).  
This periodic response of the droplet cloud persists when maintaining the same potential height but increasing its postquench width, e.g. to $\sigma=0.1 < \xi$. 
However, the spectrum of $\braket{X^2(t)}/\braket{X^2(0)}$ exhibits a slightly larger frequency, see Fig.~\ref{fig:dynamics_variance}(b). 
An outcome that is supported by our BdG predictions dictating that for fixed 
$V_0$, larger $\sigma$ values ``push'' the breathing frequency also to larger values.

Further increasing the width of the potential well, e.g. to  $\sigma=0.5 \sim \xi$ results in a comparatively more violent initial evolution and a subsequent prominent
localization. 
This can be deduced from the large amplitude of $\braket{X^2(t)}/\braket{X^2(0)}$ and its gradual decaying oscillatory pattern [characterized by a large frequency band with a small amplitude, Fig.~\ref{fig:dynamics_variance}(b)]
towards a constant value that signals the approach to a quasi-steady state at long times. 
The emergent complex response can be easily visualized in the density evolution displayed in Fig.~\ref{fig:dynamics_attractive_shallow}(c). 
Shortly after the quench, the majority of atoms are dynamically trapped by the potential well forming a central excited droplet [see also Fig.~\ref{fig:dynamics_attractive_shallow}(d)]. 
This central droplet is described to a good approximation (except from the tails) by the analytical solution in free space [Eq.~(\ref{droplet_solution})] as inferred
by the fitted waveforms illustrated in  Fig.~\ref{fig:dynamics_attractive_shallow}(d). 
The remaining atoms escape from the potential well and form
two counterpropagating density branches (matter jets) traveling to the edges, see e.g. the density profile at $t=5$ in Fig.~\ref{fig:dynamics_attractive_shallow}(d). 
Subsequently, the excited droplet undergoes a relaxation process by emitting small density fractions, reaching eventually a quasi-steady state.
Such a behavior is reminiscent of the self-evaporation mechanism observed in three-dimensional droplets~\cite{Ferioli_evaporation,fort2021self}.

\begin{figure*}[t!]
\centering
\includegraphics[width=1\linewidth]{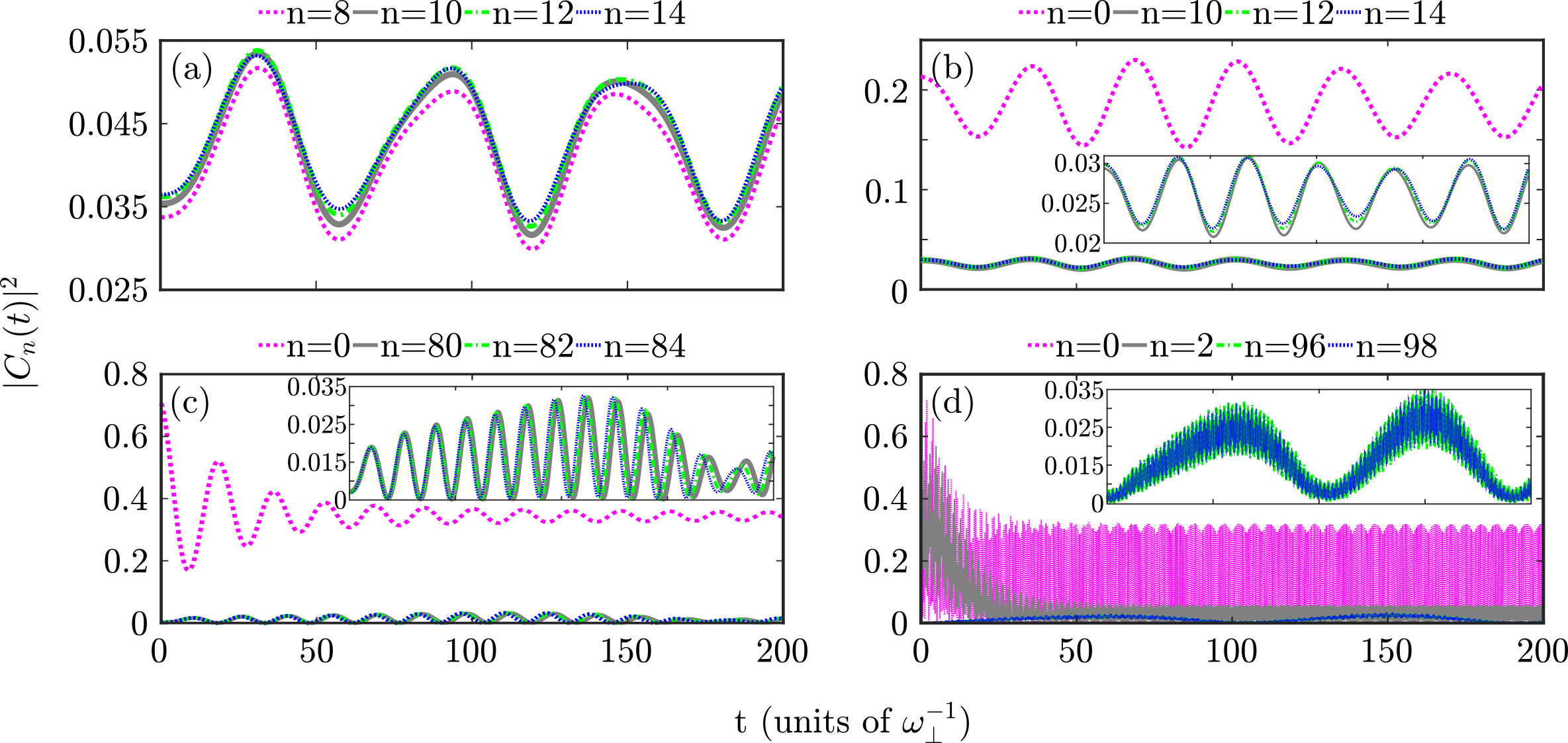}
\caption{High amplitude probabilities of finding the droplet in distinct single-particle eigenstates (see legend) of the external potential described by Eq.~(\ref{potential}) in the course of the evolution. (a) The potential barrier used has  characteristics $V_0=1$, $\sigma=0.05$, and the potential well is described by (b) $V_0=-1$, $\sigma=0.05$, (c) $V_0=-5$, $\sigma=0.05$ and (d) $V_0=-20$, $\sigma=0.5$. For a relatively shallow postquench  potential barrier or well,  several eigenstates are populated whose periodic trend is associated with the collective motion of the droplet [panels (a), (b)]. Larger potential heights lead to the occupation of a reduced number of eigenstates but of higher amplitude reflecting either the tendency to a quasi-steady state [panel (c)] or the high excitation of the droplet [panel (d)]. In all cases, the chemical potential corresponds to $\mu=-0.2$ and the LHY strength is $\delta=1$.}
\label{fig:dynamics_eigenstates}
\end{figure*} 

Turning to larger values of the postquench potential well height, e.g. $V_0=-20$, we observe a similar to the above-described (for $V_0=-1$, $\sigma=0.5 \sim \xi$) droplet response, see Fig.~\ref{fig:dynamics_attractive_strong}(a), (c).  
More concretely, independently of the width of the potential well, an initial localization takes place accompanied by the emission of two counterpropagating density segments that cannot be trapped by the potential well. 
It is, however, interesting that at least for short-times the droplet core can be  captured to a certain extent by the analytical droplet solution of Eq.~(\ref{droplet_solution}), see the fits in  Fig.~\ref{fig:dynamics_attractive_strong}(b), (d). 
Nevertheless, the emitted density segments are naturally more prominent for smaller widths, 
compare
the elevated density tails of $\abs{\psi}^2$ in Fig.~\ref{fig:dynamics_attractive_strong}(b) with the ones in panel (d). 
Following this process, a localized central droplet is produced whose excited nature can be deduced from the undulated instantaneous density profiles depicted in Fig.~\ref{fig:dynamics_attractive_strong}(b), (d). 
The degree of excitation depends also on the potential widths by means that wider ones imprint a larger amount of excitations which are in fact retained in the course of the evolution, see the vibrating density in Fig.~\ref{fig:dynamics_attractive_strong}(c) as compared to the far less disturbed one in Fig.~\ref{fig:dynamics_attractive_strong}(a).

To further inspect the dynamical relaxation tendency of the quenched droplet with respect to the height of the potential well, we present the time-evolution of the position variance, $\braket{X^2(t)}/\braket{X^2(0)}$, for different $V_0$ and fixed $\sigma$ in Fig.~\ref{fig:dynamics_variance}(c). 
It turns out that in all cases there is an initial decaying tendency of the recurring behavior of the droplet cloud at short times which is altered for quite large $V_0$ values, such as $V_0=-20$. 
For the latter quench the produced excitation is persistent enforcing internal vibrations of the droplet which performs a breathing motion. 
The characteristic frequencies of this mode are identified in the spectral peaks of the position variance shown in Fig.~\ref{fig:dynamics_variance}(d). 
These peaks clearly shift to larger frequencies for increasing $V_0$. 
According to the above description, we can infer that the characteristics of the potential well can be used not only to control the degree of droplet localization, and hence its atom number, but also its excitation frequency and dictate whether it will approach a quasi-steady state in the long time-evolution.

\subsection{Participating single-particle eigenstates}\label{SPFs}

Having described the emergent many-body evolution of the droplet within the eGPE framework a natural question that may arise is how it can be decomposed into the underlying single-particle eigenstates of the overall external potential. 
This will allow us to i) identify the competition of the underlying microscopic mechanisms, and ii) exemplify selective excitation processes. 

To achieve this, we employ the orthonormal eigenstates ($\varphi_n(x)$) of the single particle Hamiltonian, see also Sec.~\ref{sec:theory} and Fig.~\ref{fig:SP_potential}.
They serve as a basis to expand the many-body wave function, $\psi(x,t)$, obtained within the eGPE~(\ref{dimless_eGPE}) as follows $\psi(x,t)$ = $\sum_{n=1}^{N_0} C_n(t) \varphi_n(x) $. Here, $\epsilon_n$, $n=0,1,2,\dots,N_0$ denote the energy eigenvalues illustrated in Fig.~\ref{fig:SP_potential} and $N_0$ is the employed Hilbert space  truncation which depends on the quench characteristics and is chosen such that convergence of the used basis is ensured, i.e. $\sum^{N_0}_{n=1} \abs{C_n(t)}^2=1$.
For instance, a quench to ($V_0=1$, $\sigma=0.05 \ll \xi$) [($V_0=1$, $\sigma=0.5 \sim \xi$)] requires $N_0=2\times 10^3$ [$N_0=2 \times 10^3$], while using a quench to ($V_0=-1$, $\sigma=0.05$) [($V_0=-20$, $\sigma=0.5$)] necessitates $N_0= 10^3$ [$N_0=2 \times 10^3$] to affirm convergence. Importantly, $C_n(t)$ represent the expansion coefficients whose projection on the many-body wave function yields the participation probability of each individual eigenstate, namely $\abs{C_n(t)}^2=\abs{\braket{\psi(x,t)|\varphi_n(x)}}^2$. 
Apparently, if a multitude of $\abs{C_n(t)}^2$ are nonzero the resulting superposition state dictates the ensuing single-particle processes in the droplet's evolution. 
As such, the identification of the particular nonzero coefficients designates the participating eigenstates\footnote{In all discussed cases the odd eigenstates possess zero occupation due to the even parity of the initial state.} and the eigenstate transitions that are at play.

The involved single-particle eigenstates, and hence the necessity of the quantum superposition, in the course of the evolution are exemplarily shown in Fig.~\ref{fig:dynamics_eigenstates} for selected quench scenaria. 
According to our analysis, it turns out that in both the  potential barrier and well at least a few eigenstates are significantly occupied during the dynamics\footnote{Computing the quench dynamics of a single-particle wavepacket subject to the same quenches (not shown), we find substantial deviations from the discussed eGPE time-evolution evincing once more the many-body character of the discussed phenomena.}. Indeed, a sudden ramp-up of the potential barrier with $\sigma=0.05 \ll \xi$ and $V_0=1$ is accompanied by the predominant excitation of a bunch of  eigenstates mainly characterized by $n\geq 8$, see Fig.~\ref{fig:dynamics_eigenstates}(a). 
They all show a recurring behavior being a consequence of the periodic (breathing) motion of the droplet after the quench, see also Fig.~\ref{fig:dynamics_repulsive}(a). 
On the other hand, quenches to potential wells, are associated with the population of several eigenstates (whose order depends on both $V_0$ and $\sigma$) as illustrated in Figs.~\ref{fig:dynamics_eigenstates}(b)-(d) for different heights and widths of the postquench potential well. 
The oscillatory behavior of the probability amplitudes for ($V_0=-1$, $\sigma=0.05 \ll \xi$) in Fig.~\ref{fig:dynamics_eigenstates}(b) reflects the collective motion of the droplet captured, for instance, by its density distribution depicted in Fig.~\ref{fig:dynamics_attractive_shallow}(a).   
In general, quenches to larger potential heights at fixed width favor the excitation of a lesser amount of eigenstates possessing, however, higher amplitudes. 
Notice here the damped oscillation of $\abs{C_0(t)}^2$ in Fig.~\ref{fig:dynamics_eigenstates}(c) as time-evolves which essentially signifies the tendency towards a quasi-steady state of the droplet as captured also by its variance for similar parameters presented in Fig.~\ref{fig:dynamics_variance}(a). 
Finally, quenches to high potential well  amplitudes and widths facilitate the dominance of the first couple of eigenstates exhibiting high frequency oscillations, see Fig.~\ref{fig:dynamics_eigenstates}(d). 
The latter are a consequence of the substantially excited droplet core due to the quench as can be seen in the density distribution depicted in Fig.~\ref{fig:dynamics_attractive_strong}(c), (d).  
Similar features occur for the quenches to potential barriers with different characteristics (not shown).

\section{Summary \& Perspectives}\label{conclusions} 

We have investigated the existence, stability and dynamics of 1D symmetric droplet configurations under the influence of a central optical potential barrier or well.
Such potentials can be viewed as a heavy impurity interacting repulsively or attractively with their droplet host respectively.
Droplets in this context form in short-range interacting bosonic mixtures featuring intercomponent attraction but satisfying a fixed density ratio among the components dictated by their repulsive intracomponent couplings. 
This system is described by the appropriate eGPE framework encompassing the competition of attractive first-order quantum fluctuations and repulsive average mean-field interactions. 

Focusing on the ground-state of the system, we find that a potential well supporting bound states enforces spatial localization of the droplet distribution. 
The degree of localization depends both on the height and width of the potential well. 
As such, for fixed width there is a threshold in terms of the potential height above which the droplet solution ceases to exist. 
This threshold decreases for larger widths or pronounced LHY strengths due to the enhanced exerted force to the droplet in the former case, and the narrower
droplet configuration in the latter situation.  
However, a potential barrier facilitates a transition from a single- to two droplet fragments and this process occurs for relatively smaller heights when considering larger widths. 
The excitation spectrum of the above-discussed configurations has also been assessed via BdG linearization analysis. 
Stability (instability) of the structurally deformed droplets subject to a potential well (barrier) is observed since the imaginary part of the spectrum is zero (finite). 
Inspecting the real part of the spectrum it is possible to monitor the behavior of the collective mode frequencies for different parametric variations.

To trigger the dynamics we start with a harmonically trapped droplet and suddenly ramp-up the potential defect. 
A potential barrier  facilitates the dynamical splitting of the original droplet into two counterpropagating fragments towards the trap edges.  
These density fragments exhibit a droplet character which is confirmed by fitting to the analytical waveform.
Simultaneously, they become excited experiencing local vibrations and emitting small density portions due to self-evaporation~\cite{petrov2015quantum}. 
It turns out that both the dynamical separation and the velocity of these traveling droplets can be controlled via the potential characteristics, a feature that is corroborated by a simplified model based on energy conservation.  
More concretely, an increasing height (width) while keeping fixed the width (height) of the potential leads to a larger velocity and inter-droplet separation. 

In contrast, switching-on a potential well produces a somewhat more complex dynamical response. 
At short times, a relatively small density portion escapes being emitted towards the trap edges, while the majority of the atoms is trapped at the center featuring an overall dynamical localization while being excited. 
The presence of collective excitations is imprinted on the distorted density profiles and their frequency is captured by the position variance whose oscillation frequency is in line with the predictions of the excitation spectrum.  
Subsequently, the excited droplet relaxes via self-evaporation mechanisms, attaining a quasi-steady state at long evolution times.

Finally, in order to exemplify the superposition
nature of the observed dynamics and identify the participating microscopic mechanisms we have computed the population of the contributing single-particle eigenstates. 
The latter are numerically obtained through exact diagonalization of the external potential and their projections to the many-body, eGPE obtained, wave function are extracted. 
It turns out that in all cases, at least a few single-particle eigenstates contribute significantly, evincing the crucial role of the quantum superposition for the discussed phenomena. 
Moreover, the dynamical behavior of the eigenstates reflects the collective motion of the perturbed droplet or signals its tendency towards a quasi-steady state at long evolution times.

There are several intriguing extensions of the present work. 
A straightforward direction would be to study the controlled fragmentation of multiple droplets, similar to the recent three-dimensional experiment of Ref.~\cite{Cavicchioli}.  
Such fragmentation could be achieved by employing a lattice potential engineered by multiple barriers, rather than interaction quenches.
Another interesting pathway is to explore the presence of beyond LHY corrections in the ensuing non-equilibrium quantum dynamics utilizing suitable \textit{ab-initio} approaches such as the ML-MCTDHX method~\cite{cao2017unified} for exploring the emergent correlation patterns and possible modifications of the droplet's regions of existence. 
Yet another interesting perspective would be to study droplet's tunneling properties by considering a magnetic field gradient providing an energy offset among the left and right sides of the potential barrier, in analogy with the tunneling mechanisms addressed in Refs.~\cite{Abdullaev_DW,Wysocki_DW}.
In addition, extending our results to 2D quantum droplets aiming to trigger radially symmetric dispersive shock-waves and understand the impact of quantum corrections as opposed to their mean-field counterparts~\cite{Hoefer_DSW} is certainly desirable. 
In such 2D settings it would be relevant to exploit time-dependent potential barriers which may emulate gravitational phenomena in atomic systems~\cite{Sparn,Lahav}. 
Considering such quench protocols in genuine two-component settings in order to engineer un-conventional breathers and Rogue wave solutions~\cite{Peregrine_exp} is another direction worth to be pursued.

\section*{Acknowledgements}

The authors would like to thank P.G. Kevrekidis, Th. Busch, T. Fogarty, J. Pelayo, and I. Englezos for fruitful discussions on the topic of droplets. 
S.I.M. acknowledges support from the Missouri
University of Science and Technology, Department of
Physics, Startup fund.

\bibliographystyle{apsrev4-1}
\bibliography{reference}

\begin{thebibliography}{77}%
\makeatletter
\providecommand \@ifxundefined [1]{%
 \@ifx{#1\undefined}
}%
\providecommand \@ifnum [1]{%
 \ifnum #1\expandafter \@firstoftwo
 \else \expandafter \@secondoftwo
 \fi
}%
\providecommand \@ifx [1]{%
 \ifx #1\expandafter \@firstoftwo
 \else \expandafter \@secondoftwo
 \fi
}%
\providecommand \natexlab [1]{#1}%
\providecommand \enquote  [1]{``#1''}%
\providecommand \bibnamefont  [1]{#1}%
\providecommand \bibfnamefont [1]{#1}%
\providecommand \citenamefont [1]{#1}%
\providecommand \href@noop [0]{\@secondoftwo}%
\providecommand \href [0]{\begingroup \@sanitize@url \@href}%
\providecommand \@href[1]{\@@startlink{#1}\@@href}%
\providecommand \@@href[1]{\endgroup#1\@@endlink}%
\providecommand \@sanitize@url [0]{\catcode `\\12\catcode `\$12\catcode `\&12\catcode `\#12\catcode `\^12\catcode `\_12\catcode `\%12\relax}%
\providecommand \@@startlink[1]{}%
\providecommand \@@endlink[0]{}%
\providecommand \url  [0]{\begingroup\@sanitize@url \@url }%
\providecommand \@url [1]{\endgroup\@href {#1}{\urlprefix }}%
\providecommand \urlprefix  [0]{URL }%
\providecommand \Eprint [0]{\href }%
\providecommand \doibase [0]{http://dx.doi.org/}%
\providecommand \selectlanguage [0]{\@gobble}%
\providecommand \bibinfo  [0]{\@secondoftwo}%
\providecommand \bibfield  [0]{\@secondoftwo}%
\providecommand \translation [1]{[#1]}%
\providecommand \BibitemOpen [0]{}%
\providecommand \bibitemStop [0]{}%
\providecommand \bibitemNoStop [0]{.\EOS\space}%
\providecommand \EOS [0]{\spacefactor3000\relax}%
\providecommand \BibitemShut  [1]{\csname bibitem#1\endcsname}%
\let\auto@bib@innerbib\@empty
\bibitem [{\citenamefont {B{\"o}ttcher}\ \emph {et~al.}(2020)\citenamefont {B{\"o}ttcher}, \citenamefont {Schmidt}, \citenamefont {Hertkorn}, \citenamefont {N~g}, \citenamefont {Graham}, \citenamefont {Guo}, \citenamefont {Langen},\ and\ \citenamefont {Pfau}}]{bottcher2020new}%
  \BibitemOpen
  \bibfield  {author} {\bibinfo {author} {\bibfnamefont {F.}~\bibnamefont {B{\"o}ttcher}}, \bibinfo {author} {\bibfnamefont {J.-N.}\ \bibnamefont {Schmidt}}, \bibinfo {author} {\bibfnamefont {J.}~\bibnamefont {Hertkorn}}, \bibinfo {author} {\bibfnamefont {K.~S.~H.}\ \bibnamefont {N~g}}, \bibinfo {author} {\bibfnamefont {S.~D.}\ \bibnamefont {Graham}}, \bibinfo {author} {\bibfnamefont {M.}~\bibnamefont {Guo}}, \bibinfo {author} {\bibfnamefont {T.}~\bibnamefont {Langen}}, \ and\ \bibinfo {author} {\bibfnamefont {T.}~\bibnamefont {Pfau}},\ }\href {https://iopscience.iop.org/article/10.1088/1361-6633/abc9ab/meta?casa_token=6MrgTEisRGMAAAAA:EgZLCK_kD0w8POUiBsJNBkH-UD3HHM2uJL8Yh_Uaxp7yai6FkLTXhy7vtUWgvVs96PUULdqBaPzi8TztrHRZeh_bNZxO} {\bibfield  {journal} {\bibinfo  {journal} {Rep. Progr. Phys.}\ }\textbf {\bibinfo {volume} {84}},\ \bibinfo {pages} {012403} (\bibinfo {year} {2020})}\BibitemShut {NoStop}%
\bibitem [{\citenamefont {Chomaz}\ \emph {et~al.}(2023)\citenamefont {Chomaz}, \citenamefont {Ferrier-Barbut}, \citenamefont {Ferlaino}, \citenamefont {Laburthe-Tolra}, \citenamefont {Lev},\ and\ \citenamefont {Pfau}}]{chomaz2022dipolar}%
  \BibitemOpen
  \bibfield  {author} {\bibinfo {author} {\bibfnamefont {L.}~\bibnamefont {Chomaz}}, \bibinfo {author} {\bibfnamefont {I.}~\bibnamefont {Ferrier-Barbut}}, \bibinfo {author} {\bibfnamefont {F.}~\bibnamefont {Ferlaino}}, \bibinfo {author} {\bibfnamefont {B.}~\bibnamefont {Laburthe-Tolra}}, \bibinfo {author} {\bibfnamefont {B.~L.}\ \bibnamefont {Lev}}, \ and\ \bibinfo {author} {\bibfnamefont {T.}~\bibnamefont {Pfau}},\ }\href {\doibase 10.1088/1361-6633/aca814} {\bibfield  {journal} {\bibinfo  {journal} {Rep. Prog. Phys.}\ }\textbf {\bibinfo {volume} {86}},\ \bibinfo {pages} {026401} (\bibinfo {year} {2023})}\BibitemShut {NoStop}%
\bibitem [{\citenamefont {Chen}\ and\ \citenamefont {Hung}(2021)}]{Chen_observation_2021}%
  \BibitemOpen
  \bibfield  {author} {\bibinfo {author} {\bibfnamefont {C.-A.}\ \bibnamefont {Chen}}\ and\ \bibinfo {author} {\bibfnamefont {C.-L.}\ \bibnamefont {Hung}},\ }\href {\doibase 10.1103/PhysRevLett.127.023604} {\bibfield  {journal} {\bibinfo  {journal} {Phys. Rev. Lett.}\ }\textbf {\bibinfo {volume} {127}},\ \bibinfo {pages} {023604} (\bibinfo {year} {2021})}\BibitemShut {NoStop}%
\bibitem [{\citenamefont {Luo}\ \emph {et~al.}(2021)\citenamefont {Luo}, \citenamefont {Pang}, \citenamefont {Liu}, \citenamefont {Li},\ and\ \citenamefont {Malomed}}]{luo2021new}%
  \BibitemOpen
  \bibfield  {author} {\bibinfo {author} {\bibfnamefont {Z.-H.}\ \bibnamefont {Luo}}, \bibinfo {author} {\bibfnamefont {W.}~\bibnamefont {Pang}}, \bibinfo {author} {\bibfnamefont {B.}~\bibnamefont {Liu}}, \bibinfo {author} {\bibfnamefont {Y.-Y.}\ \bibnamefont {Li}}, \ and\ \bibinfo {author} {\bibfnamefont {B.~A.}\ \bibnamefont {Malomed}},\ }\href {https://link.springer.com/article/10.1007/s11467-020-1020-2} {\bibfield  {journal} {\bibinfo  {journal} {Front. Phys.}\ }\textbf {\bibinfo {volume} {16}},\ \bibinfo {pages} {1} (\bibinfo {year} {2021})}\BibitemShut {NoStop}%
\bibitem [{\citenamefont {Malomed}(2021)}]{malomed2020family}%
  \BibitemOpen
  \bibfield  {author} {\bibinfo {author} {\bibfnamefont {B.~A.}\ \bibnamefont {Malomed}},\ }\href {https://arxiv.org/abs/2010.13461} {\bibfield  {journal} {\bibinfo  {journal} {Front. Phys.}\ }\textbf {\bibinfo {volume} {16}},\ \bibinfo {pages} {22504} (\bibinfo {year} {2021})}\BibitemShut {NoStop}%
\bibitem [{\citenamefont {Mistakidis}\ \emph {et~al.}(2023)\citenamefont {Mistakidis}, \citenamefont {Volosniev}, \citenamefont {Barfknecht}, \citenamefont {Fogarty}, \citenamefont {Busch}, \citenamefont {Foerster}, \citenamefont {Schmelcher},\ and\ \citenamefont {Zinner}}]{mistakidis2023few}%
  \BibitemOpen
  \bibfield  {author} {\bibinfo {author} {\bibfnamefont {S.~I.}\ \bibnamefont {Mistakidis}}, \bibinfo {author} {\bibfnamefont {A.~G.}\ \bibnamefont {Volosniev}}, \bibinfo {author} {\bibfnamefont {R.~E.}\ \bibnamefont {Barfknecht}}, \bibinfo {author} {\bibfnamefont {T.}~\bibnamefont {Fogarty}}, \bibinfo {author} {\bibfnamefont {T.}~\bibnamefont {Busch}}, \bibinfo {author} {\bibfnamefont {A.}~\bibnamefont {Foerster}}, \bibinfo {author} {\bibfnamefont {P.}~\bibnamefont {Schmelcher}}, \ and\ \bibinfo {author} {\bibfnamefont {N.~T.}\ \bibnamefont {Zinner}},\ }\href {https://www.sciencedirect.com/science/article/abs/pii/S0370157323003162} {\bibfield  {journal} {\bibinfo  {journal} {Phys. Rep.}\ }\textbf {\bibinfo {volume} {1042}},\ \bibinfo {pages} {1} (\bibinfo {year} {2023})}\BibitemShut {NoStop}%
\bibitem [{\citenamefont {Politi}\ \emph {et~al.}(2022)\citenamefont {Politi}, \citenamefont {Trautmann}, \citenamefont {Ilzh{\"o}fer}, \citenamefont {Durastante}, \citenamefont {Mark}, \citenamefont {Modugno},\ and\ \citenamefont {Ferlaino}}]{politi2022interspecies}%
  \BibitemOpen
  \bibfield  {author} {\bibinfo {author} {\bibfnamefont {C.}~\bibnamefont {Politi}}, \bibinfo {author} {\bibfnamefont {A.}~\bibnamefont {Trautmann}}, \bibinfo {author} {\bibfnamefont {P.}~\bibnamefont {Ilzh{\"o}fer}}, \bibinfo {author} {\bibfnamefont {G.}~\bibnamefont {Durastante}}, \bibinfo {author} {\bibfnamefont {M.}~\bibnamefont {Mark}}, \bibinfo {author} {\bibfnamefont {M.}~\bibnamefont {Modugno}}, \ and\ \bibinfo {author} {\bibfnamefont {F.}~\bibnamefont {Ferlaino}},\ }\href {https://journals.aps.org/pra/abstract/10.1103/PhysRevA.105.023304} {\bibfield  {journal} {\bibinfo  {journal} {Phys. Rev. A}\ }\textbf {\bibinfo {volume} {105}},\ \bibinfo {pages} {023304} (\bibinfo {year} {2022})}\BibitemShut {NoStop}%
\bibitem [{\citenamefont {Cheiney}\ \emph {et~al.}(2018)\citenamefont {Cheiney}, \citenamefont {Cabrera}, \citenamefont {Sanz}, \citenamefont {Naylor}, \citenamefont {Tanzi},\ and\ \citenamefont {Tarruell}}]{cheiney2018bright}%
  \BibitemOpen
  \bibfield  {author} {\bibinfo {author} {\bibfnamefont {P.}~\bibnamefont {Cheiney}}, \bibinfo {author} {\bibfnamefont {C.~R.}\ \bibnamefont {Cabrera}}, \bibinfo {author} {\bibfnamefont {J.}~\bibnamefont {Sanz}}, \bibinfo {author} {\bibfnamefont {B.}~\bibnamefont {Naylor}}, \bibinfo {author} {\bibfnamefont {L.}~\bibnamefont {Tanzi}}, \ and\ \bibinfo {author} {\bibfnamefont {L.}~\bibnamefont {Tarruell}},\ }\href {https://journals.aps.org/prl/abstract/10.1103/PhysRevLett.120.135301} {\bibfield  {journal} {\bibinfo  {journal} {Phys. Rev. Lett.}\ }\textbf {\bibinfo {volume} {120}},\ \bibinfo {pages} {135301} (\bibinfo {year} {2018})}\BibitemShut {NoStop}%
\bibitem [{\citenamefont {Semeghini}\ \emph {et~al.}(2018)\citenamefont {Semeghini}, \citenamefont {Ferioli}, \citenamefont {Masi}, \citenamefont {Mazzinghi}, \citenamefont {Wolswijk}, \citenamefont {Minardi}, \citenamefont {Modugno}, \citenamefont {Modugno}, \citenamefont {Inguscio},\ and\ \citenamefont {Fattori}}]{semeghini2018self}%
  \BibitemOpen
  \bibfield  {author} {\bibinfo {author} {\bibfnamefont {G.}~\bibnamefont {Semeghini}}, \bibinfo {author} {\bibfnamefont {G.}~\bibnamefont {Ferioli}}, \bibinfo {author} {\bibfnamefont {L.}~\bibnamefont {Masi}}, \bibinfo {author} {\bibfnamefont {C.}~\bibnamefont {Mazzinghi}}, \bibinfo {author} {\bibfnamefont {L.}~\bibnamefont {Wolswijk}}, \bibinfo {author} {\bibfnamefont {F.}~\bibnamefont {Minardi}}, \bibinfo {author} {\bibfnamefont {M.}~\bibnamefont {Modugno}}, \bibinfo {author} {\bibfnamefont {G.}~\bibnamefont {Modugno}}, \bibinfo {author} {\bibfnamefont {M.}~\bibnamefont {Inguscio}}, \ and\ \bibinfo {author} {\bibfnamefont {M.}~\bibnamefont {Fattori}},\ }\href {https://journals.aps.org/prl/abstract/10.1103/PhysRevLett.120.235301} {\bibfield  {journal} {\bibinfo  {journal} {Phys. Rev. Lett.}\ }\textbf {\bibinfo {volume} {120}},\ \bibinfo {pages} {235301} (\bibinfo {year} {2018})}\BibitemShut {NoStop}%
\bibitem [{\citenamefont {Cabrera}\ \emph {et~al.}(2018)\citenamefont {Cabrera}, \citenamefont {Tanzi}, \citenamefont {Sanz}, \citenamefont {Naylor}, \citenamefont {Thomas}, \citenamefont {Cheiney},\ and\ \citenamefont {Tarruell}}]{cabrera2018quantum}%
  \BibitemOpen
  \bibfield  {author} {\bibinfo {author} {\bibfnamefont {C.~R.}\ \bibnamefont {Cabrera}}, \bibinfo {author} {\bibfnamefont {L.}~\bibnamefont {Tanzi}}, \bibinfo {author} {\bibfnamefont {J.}~\bibnamefont {Sanz}}, \bibinfo {author} {\bibfnamefont {B.}~\bibnamefont {Naylor}}, \bibinfo {author} {\bibfnamefont {P.}~\bibnamefont {Thomas}}, \bibinfo {author} {\bibfnamefont {P.}~\bibnamefont {Cheiney}}, \ and\ \bibinfo {author} {\bibfnamefont {L.}~\bibnamefont {Tarruell}},\ }\href {https://www.science.org/doi/abs/10.1126/science.aao5686} {\bibfield  {journal} {\bibinfo  {journal} {Science}\ }\textbf {\bibinfo {volume} {359}},\ \bibinfo {pages} {301} (\bibinfo {year} {2018})}\BibitemShut {NoStop}%
\bibitem [{\citenamefont {D'Errico}\ \emph {et~al.}(2019)\citenamefont {D'Errico}, \citenamefont {Burchianti}, \citenamefont {Prevedelli}, \citenamefont {Salasnich}, \citenamefont {Ancilotto}, \citenamefont {Modugno}, \citenamefont {Minardi},\ and\ \citenamefont {Fort}}]{d2019observation}%
  \BibitemOpen
  \bibfield  {author} {\bibinfo {author} {\bibfnamefont {C.}~\bibnamefont {D'Errico}}, \bibinfo {author} {\bibfnamefont {A.}~\bibnamefont {Burchianti}}, \bibinfo {author} {\bibfnamefont {M.}~\bibnamefont {Prevedelli}}, \bibinfo {author} {\bibfnamefont {L.}~\bibnamefont {Salasnich}}, \bibinfo {author} {\bibfnamefont {F.}~\bibnamefont {Ancilotto}}, \bibinfo {author} {\bibfnamefont {M.}~\bibnamefont {Modugno}}, \bibinfo {author} {\bibfnamefont {F.}~\bibnamefont {Minardi}}, \ and\ \bibinfo {author} {\bibfnamefont {C.}~\bibnamefont {Fort}},\ }\href {https://journals.aps.org/prresearch/abstract/10.1103/PhysRevResearch.1.033155} {\bibfield  {journal} {\bibinfo  {journal} {Phys. Rev. Research}\ }\textbf {\bibinfo {volume} {1}},\ \bibinfo {pages} {033155} (\bibinfo {year} {2019})}\BibitemShut {NoStop}%
\bibitem [{\citenamefont {Burchianti}\ \emph {et~al.}(2020)\citenamefont {Burchianti}, \citenamefont {D'Errico}, \citenamefont {Prevedelli}, \citenamefont {Salasnich}, \citenamefont {Ancilotto}, \citenamefont {Modugno}, \citenamefont {Minardi},\ and\ \citenamefont {Fort}}]{burchianti2020dual}%
  \BibitemOpen
  \bibfield  {author} {\bibinfo {author} {\bibfnamefont {A.}~\bibnamefont {Burchianti}}, \bibinfo {author} {\bibfnamefont {C.}~\bibnamefont {D'Errico}}, \bibinfo {author} {\bibfnamefont {M.}~\bibnamefont {Prevedelli}}, \bibinfo {author} {\bibfnamefont {L.}~\bibnamefont {Salasnich}}, \bibinfo {author} {\bibfnamefont {F.}~\bibnamefont {Ancilotto}}, \bibinfo {author} {\bibfnamefont {M.}~\bibnamefont {Modugno}}, \bibinfo {author} {\bibfnamefont {F.}~\bibnamefont {Minardi}}, \ and\ \bibinfo {author} {\bibfnamefont {C.}~\bibnamefont {Fort}},\ }\href {https://www.mdpi.com/2410-3896/5/1/21} {\bibfield  {journal} {\bibinfo  {journal} {Condens. Matter}\ }\textbf {\bibinfo {volume} {5}},\ \bibinfo {pages} {21} (\bibinfo {year} {2020})}\BibitemShut {NoStop}%
\bibitem [{\citenamefont {Ilg}\ \emph {et~al.}(2018)\citenamefont {Ilg}, \citenamefont {Kumlin}, \citenamefont {Santos}, \citenamefont {Petrov},\ and\ \citenamefont {B\"uchler}}]{Ilg_crossover_2018}%
  \BibitemOpen
  \bibfield  {author} {\bibinfo {author} {\bibfnamefont {T.}~\bibnamefont {Ilg}}, \bibinfo {author} {\bibfnamefont {J.}~\bibnamefont {Kumlin}}, \bibinfo {author} {\bibfnamefont {L.}~\bibnamefont {Santos}}, \bibinfo {author} {\bibfnamefont {D.~S.}\ \bibnamefont {Petrov}}, \ and\ \bibinfo {author} {\bibfnamefont {H.~P.}\ \bibnamefont {B\"uchler}},\ }\href {\doibase 10.1103/PhysRevA.98.051604} {\bibfield  {journal} {\bibinfo  {journal} {Phys. Rev. A}\ }\textbf {\bibinfo {volume} {98}},\ \bibinfo {pages} {051604} (\bibinfo {year} {2018})}\BibitemShut {NoStop}%
\bibitem [{\citenamefont {Pelayo}\ \emph {et~al.}(2025{\natexlab{a}})\citenamefont {Pelayo}, \citenamefont {Bougas}, \citenamefont {Fogarty}, \citenamefont {Busch},\ and\ \citenamefont {Mistakidis}}]{pelayo2025phases}%
  \BibitemOpen
  \bibfield  {author} {\bibinfo {author} {\bibfnamefont {J.~C.}\ \bibnamefont {Pelayo}}, \bibinfo {author} {\bibfnamefont {G.}~\bibnamefont {Bougas}}, \bibinfo {author} {\bibfnamefont {T.}~\bibnamefont {Fogarty}}, \bibinfo {author} {\bibfnamefont {T.}~\bibnamefont {Busch}}, \ and\ \bibinfo {author} {\bibfnamefont {S.~I.}\ \bibnamefont {Mistakidis}},\ }\href {https://www.scipost.org/10.21468/SciPostPhys.18.4.129} {\bibfield  {journal} {\bibinfo  {journal} {SciPost Physics}\ }\textbf {\bibinfo {volume} {18}},\ \bibinfo {pages} {129} (\bibinfo {year} {2025}{\natexlab{a}})}\BibitemShut {NoStop}%
\bibitem [{\citenamefont {Lee}\ \emph {et~al.}(1957)\citenamefont {Lee}, \citenamefont {Huang},\ and\ \citenamefont {Yang}}]{lee1957eigenvalues}%
  \BibitemOpen
  \bibfield  {author} {\bibinfo {author} {\bibfnamefont {T.~D.}\ \bibnamefont {Lee}}, \bibinfo {author} {\bibfnamefont {K.}~\bibnamefont {Huang}}, \ and\ \bibinfo {author} {\bibfnamefont {C.~N.}\ \bibnamefont {Yang}},\ }\href {https://journals.aps.org/pr/abstract/10.1103/PhysRev.106.1135} {\bibfield  {journal} {\bibinfo  {journal} {Phys. Rev.}\ }\textbf {\bibinfo {volume} {106}},\ \bibinfo {pages} {1135} (\bibinfo {year} {1957})}\BibitemShut {NoStop}%
\bibitem [{\citenamefont {Petrov}(2015)}]{petrov2015quantum}%
  \BibitemOpen
  \bibfield  {author} {\bibinfo {author} {\bibfnamefont {D.~S.}\ \bibnamefont {Petrov}},\ }\href {https://journals.aps.org/prl/abstract/10.1103/PhysRevLett.115.155302} {\bibfield  {journal} {\bibinfo  {journal} {Phys. Rev. Lett.}\ }\textbf {\bibinfo {volume} {115}},\ \bibinfo {pages} {155302} (\bibinfo {year} {2015})}\BibitemShut {NoStop}%
\bibitem [{\citenamefont {Petrov}\ and\ \citenamefont {Astrakharchik}(2016)}]{petrov2016ultradilute}%
  \BibitemOpen
  \bibfield  {author} {\bibinfo {author} {\bibfnamefont {D.~S.}\ \bibnamefont {Petrov}}\ and\ \bibinfo {author} {\bibfnamefont {G.~E.}\ \bibnamefont {Astrakharchik}},\ }\href {https://journals.aps.org/prl/abstract/10.1103/PhysRevLett.117.100401} {\bibfield  {journal} {\bibinfo  {journal} {Phys. Rev. Lett.}\ }\textbf {\bibinfo {volume} {117}},\ \bibinfo {pages} {100401} (\bibinfo {year} {2016})}\BibitemShut {NoStop}%
\bibitem [{\citenamefont {Englezos}\ \emph {et~al.}(2025)\citenamefont {Englezos}, \citenamefont {Schmelcher},\ and\ \citenamefont {Mistakidis}}]{englezos2025multicomponent}%
  \BibitemOpen
  \bibfield  {author} {\bibinfo {author} {\bibfnamefont {I.~A.}\ \bibnamefont {Englezos}}, \bibinfo {author} {\bibfnamefont {P.}~\bibnamefont {Schmelcher}}, \ and\ \bibinfo {author} {\bibfnamefont {S.~I.}\ \bibnamefont {Mistakidis}},\ }\href {https://arxiv.org/abs/2502.08392} {\bibfield  {journal} {\bibinfo  {journal} {arXiv:2502.08392}\ } (\bibinfo {year} {2025})}\BibitemShut {NoStop}%
\bibitem [{\citenamefont {Parisi}\ \emph {et~al.}(2019)\citenamefont {Parisi}, \citenamefont {Astrakharchik},\ and\ \citenamefont {Giorgini}}]{parisi2019liquid}%
  \BibitemOpen
  \bibfield  {author} {\bibinfo {author} {\bibfnamefont {L.}~\bibnamefont {Parisi}}, \bibinfo {author} {\bibfnamefont {G.~E.}\ \bibnamefont {Astrakharchik}}, \ and\ \bibinfo {author} {\bibfnamefont {S.}~\bibnamefont {Giorgini}},\ }\href {https://journals.aps.org/prl/abstract/10.1103/PhysRevLett.122.105302} {\bibfield  {journal} {\bibinfo  {journal} {Phys. Rev. Lett.}\ }\textbf {\bibinfo {volume} {122}},\ \bibinfo {pages} {105302} (\bibinfo {year} {2019})}\BibitemShut {NoStop}%
\bibitem [{\citenamefont {Mistakidis}\ \emph {et~al.}(2021)\citenamefont {Mistakidis}, \citenamefont {Mithun}, \citenamefont {Kevrekidis}, \citenamefont {Sadeghpour},\ and\ \citenamefont {Schmelcher}}]{mistakidis2021formation}%
  \BibitemOpen
  \bibfield  {author} {\bibinfo {author} {\bibfnamefont {S.~I.}\ \bibnamefont {Mistakidis}}, \bibinfo {author} {\bibfnamefont {T.}~\bibnamefont {Mithun}}, \bibinfo {author} {\bibfnamefont {P.~G.}\ \bibnamefont {Kevrekidis}}, \bibinfo {author} {\bibfnamefont {H.~R.}\ \bibnamefont {Sadeghpour}}, \ and\ \bibinfo {author} {\bibfnamefont {P.}~\bibnamefont {Schmelcher}},\ }\href {https://journals.aps.org/prresearch/abstract/10.1103/PhysRevResearch.3.043128} {\bibfield  {journal} {\bibinfo  {journal} {Phys. Rev. Research}\ }\textbf {\bibinfo {volume} {3}},\ \bibinfo {pages} {043128} (\bibinfo {year} {2021})}\BibitemShut {NoStop}%
\bibitem [{\citenamefont {Englezos}\ \emph {et~al.}(2023)\citenamefont {Englezos}, \citenamefont {Mistakidis},\ and\ \citenamefont {Schmelcher}}]{englezos2023correlated}%
  \BibitemOpen
  \bibfield  {author} {\bibinfo {author} {\bibfnamefont {I.~A.}\ \bibnamefont {Englezos}}, \bibinfo {author} {\bibfnamefont {S.~I.}\ \bibnamefont {Mistakidis}}, \ and\ \bibinfo {author} {\bibfnamefont {P.}~\bibnamefont {Schmelcher}},\ }\href {https://journals.aps.org/pra/abstract/10.1103/PhysRevA.107.023320} {\bibfield  {journal} {\bibinfo  {journal} {Phys. Rev. A}\ }\textbf {\bibinfo {volume} {107}},\ \bibinfo {pages} {023320} (\bibinfo {year} {2023})}\BibitemShut {NoStop}%
\bibitem [{\citenamefont {Tylutki}\ \emph {et~al.}(2020)\citenamefont {Tylutki}, \citenamefont {Astrakharchik}, \citenamefont {Malomed},\ and\ \citenamefont {Petrov}}]{tylutki2020collective}%
  \BibitemOpen
  \bibfield  {author} {\bibinfo {author} {\bibfnamefont {M.}~\bibnamefont {Tylutki}}, \bibinfo {author} {\bibfnamefont {G.~E.}\ \bibnamefont {Astrakharchik}}, \bibinfo {author} {\bibfnamefont {B.~A.}\ \bibnamefont {Malomed}}, \ and\ \bibinfo {author} {\bibfnamefont {D.~S.}\ \bibnamefont {Petrov}},\ }\href {https://journals.aps.org/pra/abstract/10.1103/PhysRevA.101.051601} {\bibfield  {journal} {\bibinfo  {journal} {Phys. Rev. A}\ }\textbf {\bibinfo {volume} {101}},\ \bibinfo {pages} {051601(R)} (\bibinfo {year} {2020})}\BibitemShut {NoStop}%
\bibitem [{\citenamefont {Ancilotto}\ \emph {et~al.}(2018)\citenamefont {Ancilotto}, \citenamefont {Barranco}, \citenamefont {Guilleumas},\ and\ \citenamefont {Pi}}]{Ancilotto_tension}%
  \BibitemOpen
  \bibfield  {author} {\bibinfo {author} {\bibfnamefont {F.}~\bibnamefont {Ancilotto}}, \bibinfo {author} {\bibfnamefont {M.}~\bibnamefont {Barranco}}, \bibinfo {author} {\bibfnamefont {M.}~\bibnamefont {Guilleumas}}, \ and\ \bibinfo {author} {\bibfnamefont {M.}~\bibnamefont {Pi}},\ }\href {\doibase 10.1103/PhysRevA.98.053623} {\bibfield  {journal} {\bibinfo  {journal} {Phys. Rev. A}\ }\textbf {\bibinfo {volume} {98}},\ \bibinfo {pages} {053623} (\bibinfo {year} {2018})}\BibitemShut {NoStop}%
\bibitem [{\citenamefont {Flynn}\ \emph {et~al.}(2024)\citenamefont {Flynn}, \citenamefont {Keepfer}, \citenamefont {Parker},\ and\ \citenamefont {Billam}}]{Flynn_harm}%
  \BibitemOpen
  \bibfield  {author} {\bibinfo {author} {\bibfnamefont {T.~A.}\ \bibnamefont {Flynn}}, \bibinfo {author} {\bibfnamefont {N.~A.}\ \bibnamefont {Keepfer}}, \bibinfo {author} {\bibfnamefont {N.~G.}\ \bibnamefont {Parker}}, \ and\ \bibinfo {author} {\bibfnamefont {T.~P.}\ \bibnamefont {Billam}},\ }\href {\doibase 10.1103/PhysRevResearch.6.013209} {\bibfield  {journal} {\bibinfo  {journal} {Phys. Rev. Res.}\ }\textbf {\bibinfo {volume} {6}},\ \bibinfo {pages} {013209} (\bibinfo {year} {2024})}\BibitemShut {NoStop}%
\bibitem [{\citenamefont {Pelayo}\ \emph {et~al.}(2025{\natexlab{b}})\citenamefont {Pelayo}, \citenamefont {Bougas}, \citenamefont {Fogarty}, \citenamefont {Busch},\ and\ \citenamefont {Mistakidis}}]{pelayo2025droplet}%
  \BibitemOpen
  \bibfield  {author} {\bibinfo {author} {\bibfnamefont {J.~C.}\ \bibnamefont {Pelayo}}, \bibinfo {author} {\bibfnamefont {G.~A.}\ \bibnamefont {Bougas}}, \bibinfo {author} {\bibfnamefont {T.}~\bibnamefont {Fogarty}}, \bibinfo {author} {\bibfnamefont {T.}~\bibnamefont {Busch}}, \ and\ \bibinfo {author} {\bibfnamefont {S.~I.}\ \bibnamefont {Mistakidis}},\ }\href@noop {} {\bibfield  {journal} {\bibinfo  {journal} {arXiv preprint arXiv:2506.09314}\ } (\bibinfo {year} {2025}{\natexlab{b}})}\BibitemShut {NoStop}%
\bibitem [{\citenamefont {Ferioli}\ \emph {et~al.}(2020)\citenamefont {Ferioli}, \citenamefont {Semeghini}, \citenamefont {Terradas-Brians\'o}, \citenamefont {Masi}, \citenamefont {Fattori},\ and\ \citenamefont {Modugno}}]{Ferioli_evaporation}%
  \BibitemOpen
  \bibfield  {author} {\bibinfo {author} {\bibfnamefont {G.}~\bibnamefont {Ferioli}}, \bibinfo {author} {\bibfnamefont {G.}~\bibnamefont {Semeghini}}, \bibinfo {author} {\bibfnamefont {S.}~\bibnamefont {Terradas-Brians\'o}}, \bibinfo {author} {\bibfnamefont {L.}~\bibnamefont {Masi}}, \bibinfo {author} {\bibfnamefont {M.}~\bibnamefont {Fattori}}, \ and\ \bibinfo {author} {\bibfnamefont {M.}~\bibnamefont {Modugno}},\ }\href {\doibase 10.1103/PhysRevResearch.2.013269} {\bibfield  {journal} {\bibinfo  {journal} {Phys. Rev. Res.}\ }\textbf {\bibinfo {volume} {2}},\ \bibinfo {pages} {013269} (\bibinfo {year} {2020})}\BibitemShut {NoStop}%
\bibitem [{\citenamefont {Fort}\ and\ \citenamefont {Modugno}(2021)}]{fort2021self}%
  \BibitemOpen
  \bibfield  {author} {\bibinfo {author} {\bibfnamefont {C.}~\bibnamefont {Fort}}\ and\ \bibinfo {author} {\bibfnamefont {M.}~\bibnamefont {Modugno}},\ }\href {https://www.mdpi.com/2076-3417/11/2/866} {\bibfield  {journal} {\bibinfo  {journal} {Applied Sciences}\ }\textbf {\bibinfo {volume} {11}},\ \bibinfo {pages} {866} (\bibinfo {year} {2021})}\BibitemShut {NoStop}%
\bibitem [{\citenamefont {Edmonds}(2023)}]{edmonds2023dark}%
  \BibitemOpen
  \bibfield  {author} {\bibinfo {author} {\bibfnamefont {M.}~\bibnamefont {Edmonds}},\ }\href {\doibase 10.1103/PhysRevResearch.5.023175} {\bibfield  {journal} {\bibinfo  {journal} {Phys. Rev. Res.}\ }\textbf {\bibinfo {volume} {5}},\ \bibinfo {pages} {023175} (\bibinfo {year} {2023})}\BibitemShut {NoStop}%
\bibitem [{\citenamefont {Katsimiga}\ \emph {et~al.}(2023{\natexlab{a}})\citenamefont {Katsimiga}, \citenamefont {Mistakidis}, \citenamefont {Koutsokostas}, \citenamefont {Frantzeskakis}, \citenamefont {Carretero-Gonz{\'a}lez},\ and\ \citenamefont {Kevrekidis}}]{katsimiga2023solitary}%
  \BibitemOpen
  \bibfield  {author} {\bibinfo {author} {\bibfnamefont {G.~C.}\ \bibnamefont {Katsimiga}}, \bibinfo {author} {\bibfnamefont {S.~I.}\ \bibnamefont {Mistakidis}}, \bibinfo {author} {\bibfnamefont {G.~N.}\ \bibnamefont {Koutsokostas}}, \bibinfo {author} {\bibfnamefont {D.~J.}\ \bibnamefont {Frantzeskakis}}, \bibinfo {author} {\bibfnamefont {R.}~\bibnamefont {Carretero-Gonz{\'a}lez}}, \ and\ \bibinfo {author} {\bibfnamefont {P.~G.}\ \bibnamefont {Kevrekidis}},\ }\href {https://journals.aps.org/pra/abstract/10.1103/PhysRevA.107.063308} {\bibfield  {journal} {\bibinfo  {journal} {Phys. Rev. A}\ }\textbf {\bibinfo {volume} {107}},\ \bibinfo {pages} {063308} (\bibinfo {year} {2023}{\natexlab{a}})}\BibitemShut {NoStop}%
\bibitem [{\citenamefont {Tengstrand}\ \emph {et~al.}(2019)\citenamefont {Tengstrand}, \citenamefont {St{\"u}rmer}, \citenamefont {Karabulut},\ and\ \citenamefont {Reimann}}]{tengstrand2019rotating}%
  \BibitemOpen
  \bibfield  {author} {\bibinfo {author} {\bibfnamefont {M.~N.}\ \bibnamefont {Tengstrand}}, \bibinfo {author} {\bibfnamefont {P.}~\bibnamefont {St{\"u}rmer}}, \bibinfo {author} {\bibfnamefont {E.~{\"O}.}\ \bibnamefont {Karabulut}}, \ and\ \bibinfo {author} {\bibfnamefont {S.~M.}\ \bibnamefont {Reimann}},\ }\href {https://journals.aps.org/prl/abstract/10.1103/PhysRevLett.123.160405} {\bibfield  {journal} {\bibinfo  {journal} {Phys. Rev. Lett.}\ }\textbf {\bibinfo {volume} {123}},\ \bibinfo {pages} {160405} (\bibinfo {year} {2019})}\BibitemShut {NoStop}%
\bibitem [{\citenamefont {Li}\ \emph {et~al.}(2018)\citenamefont {Li}, \citenamefont {Chen}, \citenamefont {Luo}, \citenamefont {Huang}, \citenamefont {Tan}, \citenamefont {Pang},\ and\ \citenamefont {Malomed}}]{li2018two}%
  \BibitemOpen
  \bibfield  {author} {\bibinfo {author} {\bibfnamefont {Y.}~\bibnamefont {Li}}, \bibinfo {author} {\bibfnamefont {Z.}~\bibnamefont {Chen}}, \bibinfo {author} {\bibfnamefont {Z.}~\bibnamefont {Luo}}, \bibinfo {author} {\bibfnamefont {C.}~\bibnamefont {Huang}}, \bibinfo {author} {\bibfnamefont {H.}~\bibnamefont {Tan}}, \bibinfo {author} {\bibfnamefont {W.}~\bibnamefont {Pang}}, \ and\ \bibinfo {author} {\bibfnamefont {B.~A.}\ \bibnamefont {Malomed}},\ }\href {https://journals.aps.org/pra/abstract/10.1103/PhysRevA.98.063602} {\bibfield  {journal} {\bibinfo  {journal} {Phys. Rev. A}\ }\textbf {\bibinfo {volume} {98}},\ \bibinfo {pages} {063602} (\bibinfo {year} {2018})}\BibitemShut {NoStop}%
\bibitem [{\citenamefont {Bougas}\ \emph {et~al.}(2024)\citenamefont {Bougas}, \citenamefont {Katsimiga}, \citenamefont {Kevrekidis},\ and\ \citenamefont {Mistakidis}}]{Bougas_vortex_drops}%
  \BibitemOpen
  \bibfield  {author} {\bibinfo {author} {\bibfnamefont {G.}~\bibnamefont {Bougas}}, \bibinfo {author} {\bibfnamefont {G.~C.}\ \bibnamefont {Katsimiga}}, \bibinfo {author} {\bibfnamefont {P.~G.}\ \bibnamefont {Kevrekidis}}, \ and\ \bibinfo {author} {\bibfnamefont {S.~I.}\ \bibnamefont {Mistakidis}},\ }\href {\doibase 10.1103/PhysRevA.110.033317} {\bibfield  {journal} {\bibinfo  {journal} {Phys. Rev. A}\ }\textbf {\bibinfo {volume} {110}},\ \bibinfo {pages} {033317} (\bibinfo {year} {2024})}\BibitemShut {NoStop}%
\bibitem [{\citenamefont {Mistakidis}\ \emph {et~al.}(2025)\citenamefont {Mistakidis}, \citenamefont {Bougas}, \citenamefont {Katsimiga},\ and\ \citenamefont {Kevrekidis}}]{Mistakidis_kinkdrop}%
  \BibitemOpen
  \bibfield  {author} {\bibinfo {author} {\bibfnamefont {S.~I.}\ \bibnamefont {Mistakidis}}, \bibinfo {author} {\bibfnamefont {G.}~\bibnamefont {Bougas}}, \bibinfo {author} {\bibfnamefont {G.~C.}\ \bibnamefont {Katsimiga}}, \ and\ \bibinfo {author} {\bibfnamefont {P.~G.}\ \bibnamefont {Kevrekidis}},\ }\href {\doibase 10.1103/PhysRevLett.134.123402} {\bibfield  {journal} {\bibinfo  {journal} {Phys. Rev. Lett.}\ }\textbf {\bibinfo {volume} {134}},\ \bibinfo {pages} {123402} (\bibinfo {year} {2025})}\BibitemShut {NoStop}%
\bibitem [{\citenamefont {Ferioli}\ \emph {et~al.}(2019)\citenamefont {Ferioli}, \citenamefont {Semeghini}, \citenamefont {Masi}, \citenamefont {Giusti}, \citenamefont {Modugno}, \citenamefont {Inguscio}, \citenamefont {Gallem{\'\i}}, \citenamefont {Recati},\ and\ \citenamefont {Fattori}}]{ferioli2019collisions}%
  \BibitemOpen
  \bibfield  {author} {\bibinfo {author} {\bibfnamefont {G.}~\bibnamefont {Ferioli}}, \bibinfo {author} {\bibfnamefont {G.}~\bibnamefont {Semeghini}}, \bibinfo {author} {\bibfnamefont {L.}~\bibnamefont {Masi}}, \bibinfo {author} {\bibfnamefont {G.}~\bibnamefont {Giusti}}, \bibinfo {author} {\bibfnamefont {G.}~\bibnamefont {Modugno}}, \bibinfo {author} {\bibfnamefont {M.}~\bibnamefont {Inguscio}}, \bibinfo {author} {\bibfnamefont {A.}~\bibnamefont {Gallem{\'\i}}}, \bibinfo {author} {\bibfnamefont {A.}~\bibnamefont {Recati}}, \ and\ \bibinfo {author} {\bibfnamefont {M.}~\bibnamefont {Fattori}},\ }\href {https://journals.aps.org/prl/abstract/10.1103/PhysRevLett.122.090401} {\bibfield  {journal} {\bibinfo  {journal} {Phys. Rev. Lett.}\ }\textbf {\bibinfo {volume} {122}},\ \bibinfo {pages} {090401} (\bibinfo {year} {2019})}\BibitemShut {NoStop}%
\bibitem [{\citenamefont {Mithun}\ \emph {et~al.}(2020)\citenamefont {Mithun}, \citenamefont {Maluckov}, \citenamefont {Kasamatsu}, \citenamefont {Malomed},\ and\ \citenamefont {Khare}}]{mithun2020modulational}%
  \BibitemOpen
  \bibfield  {author} {\bibinfo {author} {\bibfnamefont {T.}~\bibnamefont {Mithun}}, \bibinfo {author} {\bibfnamefont {A.}~\bibnamefont {Maluckov}}, \bibinfo {author} {\bibfnamefont {K.}~\bibnamefont {Kasamatsu}}, \bibinfo {author} {\bibfnamefont {B.~A.}\ \bibnamefont {Malomed}}, \ and\ \bibinfo {author} {\bibfnamefont {A.}~\bibnamefont {Khare}},\ }\href {https://www.mdpi.com/2073-8994/12/1/174} {\bibfield  {journal} {\bibinfo  {journal} {Symmetry}\ }\textbf {\bibinfo {volume} {12}},\ \bibinfo {pages} {174} (\bibinfo {year} {2020})}\BibitemShut {NoStop}%
\bibitem [{\citenamefont {Otajonov}\ \emph {et~al.}(2022)\citenamefont {Otajonov}, \citenamefont {Tsoy},\ and\ \citenamefont {Abdullaev}}]{otajonov2022modulational}%
  \BibitemOpen
  \bibfield  {author} {\bibinfo {author} {\bibfnamefont {S.~R.}\ \bibnamefont {Otajonov}}, \bibinfo {author} {\bibfnamefont {E.~N.}\ \bibnamefont {Tsoy}}, \ and\ \bibinfo {author} {\bibfnamefont {F.~K.}\ \bibnamefont {Abdullaev}},\ }\href {https://journals.aps.org/pra/abstract/10.1103/PhysRevA.106.033309} {\bibfield  {journal} {\bibinfo  {journal} {Phys. Rev. A}\ }\textbf {\bibinfo {volume} {106}},\ \bibinfo {pages} {033309} (\bibinfo {year} {2022})}\BibitemShut {NoStop}%
\bibitem [{\citenamefont {Abdullaev}\ and\ \citenamefont {Galimzyanov}(2020)}]{abdullaev2020bosonic}%
  \BibitemOpen
  \bibfield  {author} {\bibinfo {author} {\bibfnamefont {F.~K.}\ \bibnamefont {Abdullaev}}\ and\ \bibinfo {author} {\bibfnamefont {R.}~\bibnamefont {Galimzyanov}},\ }\href {https://iopscience.iop.org/article/10.1088/1361-6455/ab9659/meta?casa_token=dQ7pKuoxb0IAAAAA:20DA_plbpOHD9RCI2X8X4-jXNkqruXAKwf-kbhO2DFSWkb1Gv4xtGQbHjBmLoZdPQ0KN6yLRDt2bXjg2mqyweCka1VLW} {\bibfield  {journal} {\bibinfo  {journal} {J. Phys. B: At., Mol. and Opt. Phys.}\ }\textbf {\bibinfo {volume} {53}},\ \bibinfo {pages} {165301} (\bibinfo {year} {2020})}\BibitemShut {NoStop}%
\bibitem [{\citenamefont {Bighin}\ \emph {et~al.}(2022)\citenamefont {Bighin}, \citenamefont {Burchianti}, \citenamefont {Minardi},\ and\ \citenamefont {Macr\`{\i}}}]{Bighin_localization}%
  \BibitemOpen
  \bibfield  {author} {\bibinfo {author} {\bibfnamefont {G.}~\bibnamefont {Bighin}}, \bibinfo {author} {\bibfnamefont {A.}~\bibnamefont {Burchianti}}, \bibinfo {author} {\bibfnamefont {F.}~\bibnamefont {Minardi}}, \ and\ \bibinfo {author} {\bibfnamefont {T.}~\bibnamefont {Macr\`{\i}}},\ }\href {\doibase 10.1103/PhysRevA.106.023301} {\bibfield  {journal} {\bibinfo  {journal} {Phys. Rev. A}\ }\textbf {\bibinfo {volume} {106}},\ \bibinfo {pages} {023301} (\bibinfo {year} {2022})}\BibitemShut {NoStop}%
\bibitem [{\citenamefont {Sinha}\ \emph {et~al.}(2023)\citenamefont {Sinha}, \citenamefont {Biswas}, \citenamefont {Santos},\ and\ \citenamefont {Sinha}}]{Sinha_imp_drop}%
  \BibitemOpen
  \bibfield  {author} {\bibinfo {author} {\bibfnamefont {S.}~\bibnamefont {Sinha}}, \bibinfo {author} {\bibfnamefont {S.}~\bibnamefont {Biswas}}, \bibinfo {author} {\bibfnamefont {L.}~\bibnamefont {Santos}}, \ and\ \bibinfo {author} {\bibfnamefont {S.}~\bibnamefont {Sinha}},\ }\href {\doibase 10.1103/PhysRevA.108.023311} {\bibfield  {journal} {\bibinfo  {journal} {Phys. Rev. A}\ }\textbf {\bibinfo {volume} {108}},\ \bibinfo {pages} {023311} (\bibinfo {year} {2023})}\BibitemShut {NoStop}%
\bibitem [{\citenamefont {Wenzel}\ \emph {et~al.}(2018)\citenamefont {Wenzel}, \citenamefont {Pfau},\ and\ \citenamefont {Ferrier-Barbut}}]{wenzel2018fermionic}%
  \BibitemOpen
  \bibfield  {author} {\bibinfo {author} {\bibfnamefont {M.}~\bibnamefont {Wenzel}}, \bibinfo {author} {\bibfnamefont {T.}~\bibnamefont {Pfau}}, \ and\ \bibinfo {author} {\bibfnamefont {I.}~\bibnamefont {Ferrier-Barbut}},\ }\href {https://iopscience.iop.org/article/10.1088/1402-4896/aadd72/meta?casa_token=7NeSBwiKKqEAAAAA:bp1h0KuDXiy0Wg3PVKXIz1OcqXs1IX-UA0TMETJDNbtptMXUkWnPAVhDTMUk--mLYCUrGE-wF2sRay96Dl5-o4DtsaAj} {\bibfield  {journal} {\bibinfo  {journal} {Phys. Scripta}\ }\textbf {\bibinfo {volume} {93}},\ \bibinfo {pages} {104004} (\bibinfo {year} {2018})}\BibitemShut {NoStop}%
\bibitem [{\citenamefont {Pelayo}\ \emph {et~al.}(2024)\citenamefont {Pelayo}, \citenamefont {Fogarty}, \citenamefont {Busch},\ and\ \citenamefont {Mistakidis}}]{pelayo_BF_drops}%
  \BibitemOpen
  \bibfield  {author} {\bibinfo {author} {\bibfnamefont {J.~C.}\ \bibnamefont {Pelayo}}, \bibinfo {author} {\bibfnamefont {T.}~\bibnamefont {Fogarty}}, \bibinfo {author} {\bibfnamefont {T.}~\bibnamefont {Busch}}, \ and\ \bibinfo {author} {\bibfnamefont {S.~I.}\ \bibnamefont {Mistakidis}},\ }\href {\doibase 10.1103/PhysRevResearch.6.033219} {\bibfield  {journal} {\bibinfo  {journal} {Phys. Rev. Res.}\ }\textbf {\bibinfo {volume} {6}},\ \bibinfo {pages} {033219} (\bibinfo {year} {2024})}\BibitemShut {NoStop}%
\bibitem [{\citenamefont {Campbell}\ \emph {et~al.}(2014)\citenamefont {Campbell}, \citenamefont {Garc\'{\i}a-March}, \citenamefont {Fogarty},\ and\ \citenamefont {Busch}}]{Campbell}%
  \BibitemOpen
  \bibfield  {author} {\bibinfo {author} {\bibfnamefont {S.}~\bibnamefont {Campbell}}, \bibinfo {author} {\bibfnamefont {M.~A.}\ \bibnamefont {Garc\'{\i}a-March}}, \bibinfo {author} {\bibfnamefont {T.}~\bibnamefont {Fogarty}}, \ and\ \bibinfo {author} {\bibfnamefont {T.}~\bibnamefont {Busch}},\ }\href {\doibase 10.1103/PhysRevA.90.013617} {\bibfield  {journal} {\bibinfo  {journal} {Phys. Rev. A}\ }\textbf {\bibinfo {volume} {90}},\ \bibinfo {pages} {013617} (\bibinfo {year} {2014})}\BibitemShut {NoStop}%
\bibitem [{\citenamefont {Brauneis}\ \emph {et~al.}(2022)\citenamefont {Brauneis}, \citenamefont {Backert}, \citenamefont {Mistakidis}, \citenamefont {Lemeshko}, \citenamefont {Hammer},\ and\ \citenamefont {Volosniev}}]{brauneis2022artificial}%
  \BibitemOpen
  \bibfield  {author} {\bibinfo {author} {\bibfnamefont {F.}~\bibnamefont {Brauneis}}, \bibinfo {author} {\bibfnamefont {T.~G.}\ \bibnamefont {Backert}}, \bibinfo {author} {\bibfnamefont {S.~I.}\ \bibnamefont {Mistakidis}}, \bibinfo {author} {\bibfnamefont {M.}~\bibnamefont {Lemeshko}}, \bibinfo {author} {\bibfnamefont {H.-W.}\ \bibnamefont {Hammer}}, \ and\ \bibinfo {author} {\bibfnamefont {A.~G.}\ \bibnamefont {Volosniev}},\ }\href {https://iopscience.iop.org/article/10.1088/1367-2630/ac78d8/meta} {\bibfield  {journal} {\bibinfo  {journal} {New J. Phys.}\ }\textbf {\bibinfo {volume} {24}},\ \bibinfo {pages} {063036} (\bibinfo {year} {2022})}\BibitemShut {NoStop}%
\bibitem [{\citenamefont {Saqlain}\ \emph {et~al.}(2023)\citenamefont {Saqlain}, \citenamefont {Mithun}, \citenamefont {Carretero-Gonz\'alez},\ and\ \citenamefont {Kevrekidis}}]{Saqlain}%
  \BibitemOpen
  \bibfield  {author} {\bibinfo {author} {\bibfnamefont {S.}~\bibnamefont {Saqlain}}, \bibinfo {author} {\bibfnamefont {T.}~\bibnamefont {Mithun}}, \bibinfo {author} {\bibfnamefont {R.}~\bibnamefont {Carretero-Gonz\'alez}}, \ and\ \bibinfo {author} {\bibfnamefont {P.~G.}\ \bibnamefont {Kevrekidis}},\ }\href {\doibase 10.1103/PhysRevA.107.033310} {\bibfield  {journal} {\bibinfo  {journal} {Phys. Rev. A}\ }\textbf {\bibinfo {volume} {107}},\ \bibinfo {pages} {033310} (\bibinfo {year} {2023})}\BibitemShut {NoStop}%
\bibitem [{\citenamefont {dos Santos}\ \emph {et~al.}(2025)\citenamefont {dos Santos}, \citenamefont {Calazans~de Brito}, \citenamefont {Gammal},\ and\ \citenamefont {Kamchatnov}}]{dos2025supersonic}%
  \BibitemOpen
  \bibfield  {author} {\bibinfo {author} {\bibfnamefont {G.~H.}\ \bibnamefont {dos Santos}}, \bibinfo {author} {\bibfnamefont {L.~F.}\ \bibnamefont {Calazans~de Brito}}, \bibinfo {author} {\bibfnamefont {A.}~\bibnamefont {Gammal}}, \ and\ \bibinfo {author} {\bibfnamefont {A.~M.}\ \bibnamefont {Kamchatnov}},\ }\href {\doibase 10.1103/mxyb-6jh1} {\bibfield  {journal} {\bibinfo  {journal} {Phys. Rev. A}\ }\textbf {\bibinfo {volume} {112}},\ \bibinfo {pages} {033318} (\bibinfo {year} {2025})}\BibitemShut {NoStop}%
\bibitem [{\citenamefont {Engels}\ and\ \citenamefont {Atherton}(2007)}]{Engels_obstacle}%
  \BibitemOpen
  \bibfield  {author} {\bibinfo {author} {\bibfnamefont {P.}~\bibnamefont {Engels}}\ and\ \bibinfo {author} {\bibfnamefont {C.}~\bibnamefont {Atherton}},\ }\href {\doibase 10.1103/PhysRevLett.99.160405} {\bibfield  {journal} {\bibinfo  {journal} {Phys. Rev. Lett.}\ }\textbf {\bibinfo {volume} {99}},\ \bibinfo {pages} {160405} (\bibinfo {year} {2007})}\BibitemShut {NoStop}%
\bibitem [{\citenamefont {Hakim}(1997)}]{Hakim}%
  \BibitemOpen
  \bibfield  {author} {\bibinfo {author} {\bibfnamefont {V.}~\bibnamefont {Hakim}},\ }\href {\doibase 10.1103/PhysRevE.55.2835} {\bibfield  {journal} {\bibinfo  {journal} {Phys. Rev. E}\ }\textbf {\bibinfo {volume} {55}},\ \bibinfo {pages} {2835} (\bibinfo {year} {1997})}\BibitemShut {NoStop}%
\bibitem [{\citenamefont {Rodrigues}\ \emph {et~al.}(2009)\citenamefont {Rodrigues}, \citenamefont {Kevrekidis}, \citenamefont {Carretero-Gonz\'alez}, \citenamefont {Frantzeskakis}, \citenamefont {Schmelcher}, \citenamefont {Alexander},\ and\ \citenamefont {Kivshar}}]{Rodrigues}%
  \BibitemOpen
  \bibfield  {author} {\bibinfo {author} {\bibfnamefont {A.~S.}\ \bibnamefont {Rodrigues}}, \bibinfo {author} {\bibfnamefont {P.~G.}\ \bibnamefont {Kevrekidis}}, \bibinfo {author} {\bibfnamefont {R.}~\bibnamefont {Carretero-Gonz\'alez}}, \bibinfo {author} {\bibfnamefont {D.~J.}\ \bibnamefont {Frantzeskakis}}, \bibinfo {author} {\bibfnamefont {P.}~\bibnamefont {Schmelcher}}, \bibinfo {author} {\bibfnamefont {T.~J.}\ \bibnamefont {Alexander}}, \ and\ \bibinfo {author} {\bibfnamefont {Y.~S.}\ \bibnamefont {Kivshar}},\ }\href {\doibase 10.1103/PhysRevA.79.043603} {\bibfield  {journal} {\bibinfo  {journal} {Phys. Rev. A}\ }\textbf {\bibinfo {volume} {79}},\ \bibinfo {pages} {043603} (\bibinfo {year} {2009})}\BibitemShut {NoStop}%
\bibitem [{\citenamefont {Katsimiga}\ \emph {et~al.}(2018)\citenamefont {Katsimiga}, \citenamefont {Mistakidis}, \citenamefont {Koutentakis}, \citenamefont {Kevrekidis},\ and\ \citenamefont {Schmelcher}}]{Katsimiga_obst}%
  \BibitemOpen
  \bibfield  {author} {\bibinfo {author} {\bibfnamefont {G.~C.}\ \bibnamefont {Katsimiga}}, \bibinfo {author} {\bibfnamefont {S.~I.}\ \bibnamefont {Mistakidis}}, \bibinfo {author} {\bibfnamefont {G.~M.}\ \bibnamefont {Koutentakis}}, \bibinfo {author} {\bibfnamefont {P.~G.}\ \bibnamefont {Kevrekidis}}, \ and\ \bibinfo {author} {\bibfnamefont {P.}~\bibnamefont {Schmelcher}},\ }\href {\doibase 10.1103/PhysRevA.98.013632} {\bibfield  {journal} {\bibinfo  {journal} {Phys. Rev. A}\ }\textbf {\bibinfo {volume} {98}},\ \bibinfo {pages} {013632} (\bibinfo {year} {2018})}\BibitemShut {NoStop}%
\bibitem [{\citenamefont {Abdullaev}\ \emph {et~al.}(2025)\citenamefont {Abdullaev}, \citenamefont {Galimzyanov},\ and\ \citenamefont {Shermakhmatov}}]{Abdullaev_DW}%
  \BibitemOpen
  \bibfield  {author} {\bibinfo {author} {\bibfnamefont {F.~K.}\ \bibnamefont {Abdullaev}}, \bibinfo {author} {\bibfnamefont {R.~M.}\ \bibnamefont {Galimzyanov}}, \ and\ \bibinfo {author} {\bibfnamefont {A.~M.}\ \bibnamefont {Shermakhmatov}},\ }\href {\doibase 10.1103/hr45-tcyb} {\bibfield  {journal} {\bibinfo  {journal} {Phys. Rev. A}\ }\textbf {\bibinfo {volume} {112}},\ \bibinfo {pages} {013306} (\bibinfo {year} {2025})}\BibitemShut {NoStop}%
\bibitem [{\citenamefont {Wysocki}\ \emph {et~al.}(2024)\citenamefont {Wysocki}, \citenamefont {Jachymski}, \citenamefont {Astrakharchik},\ and\ \citenamefont {Tylutki}}]{Wysocki_DW}%
  \BibitemOpen
  \bibfield  {author} {\bibinfo {author} {\bibfnamefont {P.}~\bibnamefont {Wysocki}}, \bibinfo {author} {\bibfnamefont {K.}~\bibnamefont {Jachymski}}, \bibinfo {author} {\bibfnamefont {G.~E.}\ \bibnamefont {Astrakharchik}}, \ and\ \bibinfo {author} {\bibfnamefont {M.}~\bibnamefont {Tylutki}},\ }\href {\doibase 10.1103/PhysRevA.110.033303} {\bibfield  {journal} {\bibinfo  {journal} {Phys. Rev. A}\ }\textbf {\bibinfo {volume} {110}},\ \bibinfo {pages} {033303} (\bibinfo {year} {2024})}\BibitemShut {NoStop}%
\bibitem [{\citenamefont {Debnath}\ \emph {et~al.}(2023)\citenamefont {Debnath}, \citenamefont {Khan},\ and\ \citenamefont {Malomed}}]{debnath2023interaction}%
  \BibitemOpen
  \bibfield  {author} {\bibinfo {author} {\bibfnamefont {A.}~\bibnamefont {Debnath}}, \bibinfo {author} {\bibfnamefont {A.}~\bibnamefont {Khan}}, \ and\ \bibinfo {author} {\bibfnamefont {B.}~\bibnamefont {Malomed}},\ }\href {https://www.sciencedirect.com/science/article/abs/pii/S1007570423003787} {\bibfield  {journal} {\bibinfo  {journal} {Commun. Nonlinear Sci. Numer. Simul.}\ }\textbf {\bibinfo {volume} {126}},\ \bibinfo {pages} {107457} (\bibinfo {year} {2023})}\BibitemShut {NoStop}%
\bibitem [{\citenamefont {Atkins}\ and\ \citenamefont {Friedman}(2011)}]{atkins2011molecular}%
  \BibitemOpen
  \bibfield  {author} {\bibinfo {author} {\bibfnamefont {P.~W.}\ \bibnamefont {Atkins}}\ and\ \bibinfo {author} {\bibfnamefont {R.~S.}\ \bibnamefont {Friedman}},\ }\href {https://books.google.com/books?hl=en&lr=&id=9k-cAQAAQBAJ&oi=fnd&pg=PR7&dq=P.+W.+Atkins+and+R.+S.+Friedman,+Molecular+quantum+mechanics+(Oxford+university+press,+2011)&ots=NW4hld1nFv&sig=xId1ivfj0-K0HFLc-fSFev6KmS8#v=onepage&q&f=false} {\emph {\bibinfo {title} {Molecular quantum mechanics}}}\ (\bibinfo  {publisher} {Oxford university press},\ \bibinfo {year} {2011})\BibitemShut {NoStop}%
\bibitem [{\citenamefont {Olshanii}(1998)}]{Olshanii_conf_ind}%
  \BibitemOpen
  \bibfield  {author} {\bibinfo {author} {\bibfnamefont {M.}~\bibnamefont {Olshanii}},\ }\href {\doibase 10.1103/PhysRevLett.81.938} {\bibfield  {journal} {\bibinfo  {journal} {Phys. Rev. Lett.}\ }\textbf {\bibinfo {volume} {81}},\ \bibinfo {pages} {938} (\bibinfo {year} {1998})}\BibitemShut {NoStop}%
\bibitem [{\citenamefont {Bergeman}\ \emph {et~al.}(2003)\citenamefont {Bergeman}, \citenamefont {Moore},\ and\ \citenamefont {Olshanii}}]{Bergeman_atom_2003}%
  \BibitemOpen
  \bibfield  {author} {\bibinfo {author} {\bibfnamefont {T.}~\bibnamefont {Bergeman}}, \bibinfo {author} {\bibfnamefont {M.~G.}\ \bibnamefont {Moore}}, \ and\ \bibinfo {author} {\bibfnamefont {M.}~\bibnamefont {Olshanii}},\ }\href {https://journals.aps.org/prl/abstract/10.1103/PhysRevLett.91.163201} {\bibfield  {journal} {\bibinfo  {journal} {Phys. Rev. Lett.}\ }\textbf {\bibinfo {volume} {91}},\ \bibinfo {pages} {163201} (\bibinfo {year} {2003})}\BibitemShut {NoStop}%
\bibitem [{\citenamefont {G\"orlitz}\ \emph {et~al.}(2001)\citenamefont {G\"orlitz}, \citenamefont {Vogels}, \citenamefont {Leanhardt}, \citenamefont {Raman}, \citenamefont {Gustavson}, \citenamefont {Abo-Shaeer}, \citenamefont {Chikkatur}, \citenamefont {Gupta}, \citenamefont {Inouye}, \citenamefont {Rosenband},\ and\ \citenamefont {Ketterle}}]{Gorlitz}%
  \BibitemOpen
  \bibfield  {author} {\bibinfo {author} {\bibfnamefont {A.}~\bibnamefont {G\"orlitz}}, \bibinfo {author} {\bibfnamefont {J.~M.}\ \bibnamefont {Vogels}}, \bibinfo {author} {\bibfnamefont {A.~E.}\ \bibnamefont {Leanhardt}}, \bibinfo {author} {\bibfnamefont {C.}~\bibnamefont {Raman}}, \bibinfo {author} {\bibfnamefont {T.~L.}\ \bibnamefont {Gustavson}}, \bibinfo {author} {\bibfnamefont {J.~R.}\ \bibnamefont {Abo-Shaeer}}, \bibinfo {author} {\bibfnamefont {A.~P.}\ \bibnamefont {Chikkatur}}, \bibinfo {author} {\bibfnamefont {S.}~\bibnamefont {Gupta}}, \bibinfo {author} {\bibfnamefont {S.}~\bibnamefont {Inouye}}, \bibinfo {author} {\bibfnamefont {T.}~\bibnamefont {Rosenband}}, \ and\ \bibinfo {author} {\bibfnamefont {W.}~\bibnamefont {Ketterle}},\ }\href {\doibase 10.1103/PhysRevLett.87.130402} {\bibfield  {journal} {\bibinfo  {journal} {Phys. Rev. Lett.}\ }\textbf {\bibinfo {volume} {87}},\ \bibinfo {pages} {130402} (\bibinfo {year} {2001})}\BibitemShut {NoStop}%
\bibitem [{\citenamefont {Moritz}\ \emph {et~al.}(2003)\citenamefont {Moritz}, \citenamefont {St\"oferle}, \citenamefont {K\"ohl},\ and\ \citenamefont {Esslinger}}]{Moritz}%
  \BibitemOpen
  \bibfield  {author} {\bibinfo {author} {\bibfnamefont {H.}~\bibnamefont {Moritz}}, \bibinfo {author} {\bibfnamefont {T.}~\bibnamefont {St\"oferle}}, \bibinfo {author} {\bibfnamefont {M.}~\bibnamefont {K\"ohl}}, \ and\ \bibinfo {author} {\bibfnamefont {T.}~\bibnamefont {Esslinger}},\ }\href {\doibase 10.1103/PhysRevLett.91.250402} {\bibfield  {journal} {\bibinfo  {journal} {Phys. Rev. Lett.}\ }\textbf {\bibinfo {volume} {91}},\ \bibinfo {pages} {250402} (\bibinfo {year} {2003})}\BibitemShut {NoStop}%
\bibitem [{\citenamefont {Meinert}\ \emph {et~al.}(2017)\citenamefont {Meinert}, \citenamefont {Knap}, \citenamefont {Kirilov}, \citenamefont {Jag-Lauber}, \citenamefont {Zvonarev}, \citenamefont {Demler},\ and\ \citenamefont {N{\"a}gerl}}]{meinert2017bloch}%
  \BibitemOpen
  \bibfield  {author} {\bibinfo {author} {\bibfnamefont {F.}~\bibnamefont {Meinert}}, \bibinfo {author} {\bibfnamefont {M.}~\bibnamefont {Knap}}, \bibinfo {author} {\bibfnamefont {E.}~\bibnamefont {Kirilov}}, \bibinfo {author} {\bibfnamefont {K.}~\bibnamefont {Jag-Lauber}}, \bibinfo {author} {\bibfnamefont {M.~B.}\ \bibnamefont {Zvonarev}}, \bibinfo {author} {\bibfnamefont {E.}~\bibnamefont {Demler}}, \ and\ \bibinfo {author} {\bibfnamefont {H.-C.}\ \bibnamefont {N{\"a}gerl}},\ }\href {https://www.science.org/doi/full/10.1126/science.aah6616} {\bibfield  {journal} {\bibinfo  {journal} {Science}\ }\textbf {\bibinfo {volume} {356}},\ \bibinfo {pages} {945} (\bibinfo {year} {2017})}\BibitemShut {NoStop}%
\bibitem [{\citenamefont {Romero-Ros}\ \emph {et~al.}(2024)\citenamefont {Romero-Ros}, \citenamefont {Katsimiga}, \citenamefont {Mistakidis}, \citenamefont {Mossman}, \citenamefont {Biondini}, \citenamefont {Schmelcher}, \citenamefont {Engels},\ and\ \citenamefont {Kevrekidis}}]{Peregrine_exp}%
  \BibitemOpen
  \bibfield  {author} {\bibinfo {author} {\bibfnamefont {A.}~\bibnamefont {Romero-Ros}}, \bibinfo {author} {\bibfnamefont {G.~C.}\ \bibnamefont {Katsimiga}}, \bibinfo {author} {\bibfnamefont {S.~I.}\ \bibnamefont {Mistakidis}}, \bibinfo {author} {\bibfnamefont {S.}~\bibnamefont {Mossman}}, \bibinfo {author} {\bibfnamefont {G.}~\bibnamefont {Biondini}}, \bibinfo {author} {\bibfnamefont {P.}~\bibnamefont {Schmelcher}}, \bibinfo {author} {\bibfnamefont {P.}~\bibnamefont {Engels}}, \ and\ \bibinfo {author} {\bibfnamefont {P.~G.}\ \bibnamefont {Kevrekidis}},\ }\href {\doibase 10.1103/PhysRevLett.132.033402} {\bibfield  {journal} {\bibinfo  {journal} {Phys. Rev. Lett.}\ }\textbf {\bibinfo {volume} {132}},\ \bibinfo {pages} {033402} (\bibinfo {year} {2024})}\BibitemShut {NoStop}%
\bibitem [{\citenamefont {Onofrio}\ \emph {et~al.}(2000)\citenamefont {Onofrio}, \citenamefont {Raman}, \citenamefont {Vogels}, \citenamefont {Abo-Shaeer}, \citenamefont {Chikkatur},\ and\ \citenamefont {Ketterle}}]{Onofrio}%
  \BibitemOpen
  \bibfield  {author} {\bibinfo {author} {\bibfnamefont {R.}~\bibnamefont {Onofrio}}, \bibinfo {author} {\bibfnamefont {C.}~\bibnamefont {Raman}}, \bibinfo {author} {\bibfnamefont {J.~M.}\ \bibnamefont {Vogels}}, \bibinfo {author} {\bibfnamefont {J.~R.}\ \bibnamefont {Abo-Shaeer}}, \bibinfo {author} {\bibfnamefont {A.~P.}\ \bibnamefont {Chikkatur}}, \ and\ \bibinfo {author} {\bibfnamefont {W.}~\bibnamefont {Ketterle}},\ }\href {\doibase 10.1103/PhysRevLett.85.2228} {\bibfield  {journal} {\bibinfo  {journal} {Phys. Rev. Lett.}\ }\textbf {\bibinfo {volume} {85}},\ \bibinfo {pages} {2228} (\bibinfo {year} {2000})}\BibitemShut {NoStop}%
\bibitem [{\citenamefont {Lim}\ \emph {et~al.}(2022)\citenamefont {Lim}, \citenamefont {Lee}, \citenamefont {Goo}, \citenamefont {Bae},\ and\ \citenamefont {Shin}}]{lim2022vortex}%
  \BibitemOpen
  \bibfield  {author} {\bibinfo {author} {\bibfnamefont {Y.}~\bibnamefont {Lim}}, \bibinfo {author} {\bibfnamefont {Y.}~\bibnamefont {Lee}}, \bibinfo {author} {\bibfnamefont {J.}~\bibnamefont {Goo}}, \bibinfo {author} {\bibfnamefont {D.}~\bibnamefont {Bae}}, \ and\ \bibinfo {author} {\bibfnamefont {Y.-i.}\ \bibnamefont {Shin}},\ }\href {https://iopscience.iop.org/article/10.1088/1367-2630/ac8656/meta} {\bibfield  {journal} {\bibinfo  {journal} {New J. Phys.}\ }\textbf {\bibinfo {volume} {24}},\ \bibinfo {pages} {083020} (\bibinfo {year} {2022})}\BibitemShut {NoStop}%
\bibitem [{\citenamefont {Cavicchioli}\ \emph {et~al.}(2025)\citenamefont {Cavicchioli}, \citenamefont {Fort}, \citenamefont {Ancilotto}, \citenamefont {Modugno}, \citenamefont {Minardi},\ and\ \citenamefont {Burchianti}}]{Cavicchioli}%
  \BibitemOpen
  \bibfield  {author} {\bibinfo {author} {\bibfnamefont {L.}~\bibnamefont {Cavicchioli}}, \bibinfo {author} {\bibfnamefont {C.}~\bibnamefont {Fort}}, \bibinfo {author} {\bibfnamefont {F.}~\bibnamefont {Ancilotto}}, \bibinfo {author} {\bibfnamefont {M.}~\bibnamefont {Modugno}}, \bibinfo {author} {\bibfnamefont {F.}~\bibnamefont {Minardi}}, \ and\ \bibinfo {author} {\bibfnamefont {A.}~\bibnamefont {Burchianti}},\ }\href {\doibase 10.1103/PhysRevLett.134.093401} {\bibfield  {journal} {\bibinfo  {journal} {Phys. Rev. Lett.}\ }\textbf {\bibinfo {volume} {134}},\ \bibinfo {pages} {093401} (\bibinfo {year} {2025})}\BibitemShut {NoStop}%
\bibitem [{\citenamefont {Lewenstein}\ \emph {et~al.}(2012)\citenamefont {Lewenstein}, \citenamefont {Sanpera},\ and\ \citenamefont {Ahufinger}}]{lewenstein2012ultracold}%
  \BibitemOpen
  \bibfield  {author} {\bibinfo {author} {\bibfnamefont {M.}~\bibnamefont {Lewenstein}}, \bibinfo {author} {\bibfnamefont {A.}~\bibnamefont {Sanpera}}, \ and\ \bibinfo {author} {\bibfnamefont {V.}~\bibnamefont {Ahufinger}},\ }\href {https://books.google.com/books?hl=en&lr=&id=e4OiwSkmtQwC&oi=fnd&pg=PP1&dq=M.+Lewenstein,+A.+Sanpera,+and+V.+Ahufinger,+Ultracold+Atoms+in+Optical+Lattices:+Simulating+quantum+many-+body+systems+(Oxford+University+Press+(UK),+2012)&ots=Nl0Yo-pWNi&sig=F_ylTmERVuRn7aZUjtvS_g4yT8M#v=onepage&q&f=false} {\emph {\bibinfo {title} {Ultracold Atoms in Optical Lattices: Simulating quantum many-body systems}}}\ (\bibinfo  {publisher} {Oxford University Press (UK)},\ \bibinfo {year} {2012})\BibitemShut {NoStop}%
\bibitem [{\citenamefont {Kelley}(2003)}]{kelley2003solving}%
  \BibitemOpen
  \bibfield  {author} {\bibinfo {author} {\bibfnamefont {C.~T.}\ \bibnamefont {Kelley}},\ }\href {https://scholar.google.com/scholar?hl=en&as_sdt=0%2C26&q=C.+T.+Kelley%2C+Solving+nonlinear+equations+with+Newton%E2%80%99s+method+%28SIAM%2C+2003%29.&btnG=} {\emph {\bibinfo {title} {Solving nonlinear equations with Newton's method}}}\ (\bibinfo  {publisher} {SIAM},\ \bibinfo {year} {2003})\BibitemShut {NoStop}%
\bibitem [{\citenamefont {Astrakharchik}\ and\ \citenamefont {Malomed}(2018)}]{astrakharchik2018dynamics}%
  \BibitemOpen
  \bibfield  {author} {\bibinfo {author} {\bibfnamefont {G.~E.}\ \bibnamefont {Astrakharchik}}\ and\ \bibinfo {author} {\bibfnamefont {B.~A.}\ \bibnamefont {Malomed}},\ }\href {https://journals.aps.org/pra/abstract/10.1103/PhysRevA.98.013631} {\bibfield  {journal} {\bibinfo  {journal} {Phys. Rev. A}\ }\textbf {\bibinfo {volume} {98}},\ \bibinfo {pages} {013631} (\bibinfo {year} {2018})}\BibitemShut {NoStop}%
\bibitem [{\citenamefont {Katsimiga}\ \emph {et~al.}(2023{\natexlab{b}})\citenamefont {Katsimiga}, \citenamefont {Mistakidis}, \citenamefont {Malomed}, \citenamefont {Frantzeskakis}, \citenamefont {Carretero-Gonzalez},\ and\ \citenamefont {Kevrekidis}}]{katsimiga2023interactions}%
  \BibitemOpen
  \bibfield  {author} {\bibinfo {author} {\bibfnamefont {G.~C.}\ \bibnamefont {Katsimiga}}, \bibinfo {author} {\bibfnamefont {S.~I.}\ \bibnamefont {Mistakidis}}, \bibinfo {author} {\bibfnamefont {B.~A.}\ \bibnamefont {Malomed}}, \bibinfo {author} {\bibfnamefont {D.~J.}\ \bibnamefont {Frantzeskakis}}, \bibinfo {author} {\bibfnamefont {R.}~\bibnamefont {Carretero-Gonzalez}}, \ and\ \bibinfo {author} {\bibfnamefont {P.~G.}\ \bibnamefont {Kevrekidis}},\ }\href {https://www.mdpi.com/2410-3896/8/3/67} {\bibfield  {journal} {\bibinfo  {journal} {Condens. Matter}\ }\textbf {\bibinfo {volume} {8}},\ \bibinfo {pages} {67} (\bibinfo {year} {2023}{\natexlab{b}})}\BibitemShut {NoStop}%
\bibitem [{\citenamefont {Stoof}\ \emph {et~al.}(2009)\citenamefont {Stoof}, \citenamefont {Gubbels},\ and\ \citenamefont {Dickerscheid}}]{stoof2009bose}%
  \BibitemOpen
  \bibfield  {author} {\bibinfo {author} {\bibfnamefont {H.~T.}\ \bibnamefont {Stoof}}, \bibinfo {author} {\bibfnamefont {K.~B.}\ \bibnamefont {Gubbels}}, \ and\ \bibinfo {author} {\bibfnamefont {D.~B.}\ \bibnamefont {Dickerscheid}},\ }\href {https://link.springer.com/book/10.1007/978-1-4020-8763-9} {\bibfield  {journal} {\bibinfo  {journal} {Ultracold Quantum Fields}\ ,\ \bibinfo {pages} {235}} (\bibinfo {year} {2009})}\BibitemShut {NoStop}%
\bibitem [{\citenamefont {Fei}\ \emph {et~al.}(2024)\citenamefont {Fei}, \citenamefont {Du}, \citenamefont {Chen},\ and\ \citenamefont {Zhang}}]{Fei_2Dspec}%
  \BibitemOpen
  \bibfield  {author} {\bibinfo {author} {\bibfnamefont {Y.}~\bibnamefont {Fei}}, \bibinfo {author} {\bibfnamefont {X.}~\bibnamefont {Du}}, \bibinfo {author} {\bibfnamefont {X.-L.}\ \bibnamefont {Chen}}, \ and\ \bibinfo {author} {\bibfnamefont {Y.}~\bibnamefont {Zhang}},\ }\href {\doibase 10.1103/PhysRevA.109.053309} {\bibfield  {journal} {\bibinfo  {journal} {Phys. Rev. A}\ }\textbf {\bibinfo {volume} {109}},\ \bibinfo {pages} {053309} (\bibinfo {year} {2024})}\BibitemShut {NoStop}%
\bibitem [{\citenamefont {Charalampidis}\ and\ \citenamefont {Mistakidis}(2025)}]{Charalampidis_2comp_drops}%
  \BibitemOpen
  \bibfield  {author} {\bibinfo {author} {\bibfnamefont {E.~G.}\ \bibnamefont {Charalampidis}}\ and\ \bibinfo {author} {\bibfnamefont {S.~I.}\ \bibnamefont {Mistakidis}},\ }\href {\doibase 10.1103/PhysRevA.111.013318} {\bibfield  {journal} {\bibinfo  {journal} {Phys. Rev. A}\ }\textbf {\bibinfo {volume} {111}},\ \bibinfo {pages} {013318} (\bibinfo {year} {2025})}\BibitemShut {NoStop}%
\bibitem [{\citenamefont {Liu}\ \emph {et~al.}(2022)\citenamefont {Liu}, \citenamefont {Chen}, \citenamefont {Yang}, \citenamefont {Cai}, \citenamefont {Liu}, \citenamefont {Luo}, \citenamefont {Qin}, \citenamefont {Da~Jiang}, \citenamefont {Li},\ and\ \citenamefont {Malomed}}]{liu2022vortex}%
  \BibitemOpen
  \bibfield  {author} {\bibinfo {author} {\bibfnamefont {B.}~\bibnamefont {Liu}}, \bibinfo {author} {\bibfnamefont {Y.~X.}\ \bibnamefont {Chen}}, \bibinfo {author} {\bibfnamefont {A.~W.}\ \bibnamefont {Yang}}, \bibinfo {author} {\bibfnamefont {X.~Y.}\ \bibnamefont {Cai}}, \bibinfo {author} {\bibfnamefont {Y.}~\bibnamefont {Liu}}, \bibinfo {author} {\bibfnamefont {Z.~H.}\ \bibnamefont {Luo}}, \bibinfo {author} {\bibfnamefont {X.~Z.}\ \bibnamefont {Qin}}, \bibinfo {author} {\bibfnamefont {X.}~\bibnamefont {Da~Jiang}}, \bibinfo {author} {\bibfnamefont {Y.~Y.}\ \bibnamefont {Li}}, \ and\ \bibinfo {author} {\bibfnamefont {B.~A.}\ \bibnamefont {Malomed}},\ }\href {https://iopscience.iop.org/article/10.1088/1367-2630/acab26/meta} {\bibfield  {journal} {\bibinfo  {journal} {New Journal of Physics}\ }\textbf {\bibinfo {volume} {24}},\ \bibinfo {pages} {123026} (\bibinfo {year} {2022})}\BibitemShut {NoStop}%
\bibitem [{\citenamefont {Grusdt}\ \emph {et~al.}(2024)\citenamefont {Grusdt}, \citenamefont {Mostaan}, \citenamefont {Demler},\ and\ \citenamefont {Pe{\~n}a~Ardila}}]{grusdt2024impurities}%
  \BibitemOpen
  \bibfield  {author} {\bibinfo {author} {\bibfnamefont {F.}~\bibnamefont {Grusdt}}, \bibinfo {author} {\bibfnamefont {N.}~\bibnamefont {Mostaan}}, \bibinfo {author} {\bibfnamefont {E.}~\bibnamefont {Demler}}, \ and\ \bibinfo {author} {\bibfnamefont {L.~A. A.~P.}\ \bibnamefont {Pe{\~n}a~Ardila}},\ }\href {https://iopscience.iop.org/article/10.1088/1361-6633/add94b/meta} {\bibfield  {journal} {\bibinfo  {journal} {Rep. Progr. Phys.}\ } (\bibinfo {year} {2024})}\BibitemShut {NoStop}%
\bibitem [{\citenamefont {Ronzheimer}\ \emph {et~al.}(2013)\citenamefont {Ronzheimer}, \citenamefont {Schreiber}, \citenamefont {Braun}, \citenamefont {Hodgman}, \citenamefont {Langer}, \citenamefont {McCulloch}, \citenamefont {Heidrich-Meisner}, \citenamefont {Bloch},\ and\ \citenamefont {Schneider}}]{Ronzheimer}%
  \BibitemOpen
  \bibfield  {author} {\bibinfo {author} {\bibfnamefont {J.~P.}\ \bibnamefont {Ronzheimer}}, \bibinfo {author} {\bibfnamefont {M.}~\bibnamefont {Schreiber}}, \bibinfo {author} {\bibfnamefont {S.}~\bibnamefont {Braun}}, \bibinfo {author} {\bibfnamefont {S.~S.}\ \bibnamefont {Hodgman}}, \bibinfo {author} {\bibfnamefont {S.}~\bibnamefont {Langer}}, \bibinfo {author} {\bibfnamefont {I.~P.}\ \bibnamefont {McCulloch}}, \bibinfo {author} {\bibfnamefont {F.}~\bibnamefont {Heidrich-Meisner}}, \bibinfo {author} {\bibfnamefont {I.}~\bibnamefont {Bloch}}, \ and\ \bibinfo {author} {\bibfnamefont {U.}~\bibnamefont {Schneider}},\ }\href {\doibase 10.1103/PhysRevLett.110.205301} {\bibfield  {journal} {\bibinfo  {journal} {Phys. Rev. Lett.}\ }\textbf {\bibinfo {volume} {110}},\ \bibinfo {pages} {205301} (\bibinfo {year} {2013})}\BibitemShut {NoStop}%
\bibitem [{\citenamefont {Mistakidis}\ \emph {et~al.}(2020)\citenamefont {Mistakidis}, \citenamefont {Koutentakis}, \citenamefont {Katsimiga}, \citenamefont {Busch},\ and\ \citenamefont {Schmelcher}}]{mistakidis2020many}%
  \BibitemOpen
  \bibfield  {author} {\bibinfo {author} {\bibfnamefont {S.~I.}\ \bibnamefont {Mistakidis}}, \bibinfo {author} {\bibfnamefont {G.}~\bibnamefont {Koutentakis}}, \bibinfo {author} {\bibfnamefont {G.}~\bibnamefont {Katsimiga}}, \bibinfo {author} {\bibfnamefont {T.}~\bibnamefont {Busch}}, \ and\ \bibinfo {author} {\bibfnamefont {P.}~\bibnamefont {Schmelcher}},\ }\href {https://iopscience.iop.org/article/10.1088/1367-2630/ab7599/meta} {\bibfield  {journal} {\bibinfo  {journal} {New J. Phys.}\ }\textbf {\bibinfo {volume} {22}},\ \bibinfo {pages} {043007} (\bibinfo {year} {2020})}\BibitemShut {NoStop}%
\bibitem [{\citenamefont {Cao}\ \emph {et~al.}(2017)\citenamefont {Cao}, \citenamefont {Bolsinger}, \citenamefont {Mistakidis}, \citenamefont {Koutentakis}, \citenamefont {Kr{\"o}nke}, \citenamefont {Schurer},\ and\ \citenamefont {Schmelcher}}]{cao2017unified}%
  \BibitemOpen
  \bibfield  {author} {\bibinfo {author} {\bibfnamefont {L.}~\bibnamefont {Cao}}, \bibinfo {author} {\bibfnamefont {V.}~\bibnamefont {Bolsinger}}, \bibinfo {author} {\bibfnamefont {S.~I.}\ \bibnamefont {Mistakidis}}, \bibinfo {author} {\bibfnamefont {G.~M.}\ \bibnamefont {Koutentakis}}, \bibinfo {author} {\bibfnamefont {S.}~\bibnamefont {Kr{\"o}nke}}, \bibinfo {author} {\bibfnamefont {J.~M.}\ \bibnamefont {Schurer}}, \ and\ \bibinfo {author} {\bibfnamefont {P.}~\bibnamefont {Schmelcher}},\ }\href {https://pubs.aip.org/aip/jcp/article-abstract/147/4/044106/154883/A-unified-ab-initio-approach-to-the-correlated?redirectedFrom=fulltext} {\bibfield  {journal} {\bibinfo  {journal} {J. Chem. Phys.}\ }\textbf {\bibinfo {volume} {147}},\ \bibinfo {pages} {044106} (\bibinfo {year} {2017})}\BibitemShut {NoStop}%
\bibitem [{\citenamefont {Hoefer}\ \emph {et~al.}(2006)\citenamefont {Hoefer}, \citenamefont {Ablowitz}, \citenamefont {Coddington}, \citenamefont {Cornell}, \citenamefont {Engels},\ and\ \citenamefont {Schweikhard}}]{Hoefer_DSW}%
  \BibitemOpen
  \bibfield  {author} {\bibinfo {author} {\bibfnamefont {M.~A.}\ \bibnamefont {Hoefer}}, \bibinfo {author} {\bibfnamefont {M.~J.}\ \bibnamefont {Ablowitz}}, \bibinfo {author} {\bibfnamefont {I.}~\bibnamefont {Coddington}}, \bibinfo {author} {\bibfnamefont {E.~A.}\ \bibnamefont {Cornell}}, \bibinfo {author} {\bibfnamefont {P.}~\bibnamefont {Engels}}, \ and\ \bibinfo {author} {\bibfnamefont {V.}~\bibnamefont {Schweikhard}},\ }\href {\doibase 10.1103/PhysRevA.74.023623} {\bibfield  {journal} {\bibinfo  {journal} {Phys. Rev. A}\ }\textbf {\bibinfo {volume} {74}},\ \bibinfo {pages} {023623} (\bibinfo {year} {2006})}\BibitemShut {NoStop}%
\bibitem [{\citenamefont {Sparn}\ \emph {et~al.}(2024)\citenamefont {Sparn}, \citenamefont {Kath}, \citenamefont {Liebster}, \citenamefont {Duchene}, \citenamefont {Schmidt}, \citenamefont {Tolosa-Sime\'on}, \citenamefont {Parra-L\'opez}, \citenamefont {Floerchinger}, \citenamefont {Strobel},\ and\ \citenamefont {Oberthaler}}]{Sparn}%
  \BibitemOpen
  \bibfield  {author} {\bibinfo {author} {\bibfnamefont {M.}~\bibnamefont {Sparn}}, \bibinfo {author} {\bibfnamefont {E.}~\bibnamefont {Kath}}, \bibinfo {author} {\bibfnamefont {N.}~\bibnamefont {Liebster}}, \bibinfo {author} {\bibfnamefont {J.}~\bibnamefont {Duchene}}, \bibinfo {author} {\bibfnamefont {C.~F.}\ \bibnamefont {Schmidt}}, \bibinfo {author} {\bibfnamefont {M.}~\bibnamefont {Tolosa-Sime\'on}}, \bibinfo {author} {\bibfnamefont {A.}~\bibnamefont {Parra-L\'opez}}, \bibinfo {author} {\bibfnamefont {S.}~\bibnamefont {Floerchinger}}, \bibinfo {author} {\bibfnamefont {H.}~\bibnamefont {Strobel}}, \ and\ \bibinfo {author} {\bibfnamefont {M.~K.}\ \bibnamefont {Oberthaler}},\ }\href {\doibase 10.1103/PhysRevLett.133.260201} {\bibfield  {journal} {\bibinfo  {journal} {Phys. Rev. Lett.}\ }\textbf {\bibinfo {volume} {133}},\ \bibinfo {pages} {260201} (\bibinfo {year} {2024})}\BibitemShut {NoStop}%
\bibitem [{\citenamefont {Lahav}\ \emph {et~al.}(2010)\citenamefont {Lahav}, \citenamefont {Itah}, \citenamefont {Blumkin}, \citenamefont {Gordon}, \citenamefont {Rinott}, \citenamefont {Zayats},\ and\ \citenamefont {Steinhauer}}]{Lahav}%
  \BibitemOpen
  \bibfield  {author} {\bibinfo {author} {\bibfnamefont {O.}~\bibnamefont {Lahav}}, \bibinfo {author} {\bibfnamefont {A.}~\bibnamefont {Itah}}, \bibinfo {author} {\bibfnamefont {A.}~\bibnamefont {Blumkin}}, \bibinfo {author} {\bibfnamefont {C.}~\bibnamefont {Gordon}}, \bibinfo {author} {\bibfnamefont {S.}~\bibnamefont {Rinott}}, \bibinfo {author} {\bibfnamefont {A.}~\bibnamefont {Zayats}}, \ and\ \bibinfo {author} {\bibfnamefont {J.}~\bibnamefont {Steinhauer}},\ }\href {\doibase 10.1103/PhysRevLett.105.240401} {\bibfield  {journal} {\bibinfo  {journal} {Phys. Rev. Lett.}\ }\textbf {\bibinfo {volume} {105}},\ \bibinfo {pages} {240401} (\bibinfo {year} {2010})}\BibitemShut {NoStop}%
\end{thebibliography}%

\end{document}